\documentclass{article}

\usepackage{arxiv}

\usepackage[utf8]{inputenc} 
\usepackage[T1]{fontenc}    
\usepackage{hyperref}       
\usepackage{url}            
\usepackage{booktabs}       
\usepackage{amsfonts}       
\usepackage{nicefrac}       
\usepackage{microtype}      
\usepackage{graphicx}
\usepackage{amsmath,amssymb,dsfont,array}
\usepackage{multirow, multicol, longtable, lscape, caption}

\title{Weighted Cox regression for the prediction of heterogeneous patient subgroups}

\author{
  Katrin~Madjar \\
  Department of Statistics\\
	TU Dortmund University\\
	44221 Dortmund, Germany \\
  \texttt{madjar@statistik.tu-dortmund.de} \\
   \And
 Jörg~Rahnenführer \\
 Department of Statistics\\
	TU Dortmund University\\
	44221 Dortmund, Germany \\
  \texttt{rahnenfuehrer@statistik.tu-dortmund.de} \\
}

\makeatletter
\g@addto@macro\appendix{\setcounter{figure}{0}}
\g@addto@macro\appendix{\setcounter{table}{0}}
\makeatother

\begin{document}
\maketitle

\begin{abstract}
An important task in clinical medicine is the construction of risk prediction models for specific subgroups of patients based on high-dimensional molecular measurements such as gene expression data. 
Major objectives in modeling high-dimensional data are good prediction performance and feature selection to find a subset of predictors that are truly associated with a clinical outcome such as a time-to-event endpoint.
In clinical practice, this task is challenging since patient cohorts are typically small and can be heterogeneous with regard to their relationship between predictors and outcome.
When data of several subgroups of patients with the same or similar disease are available, it is tempting to combine them to increase sample size, such as in multicenter studies. 
However, heterogeneity between subgroups can lead to biased results and subgroup-specific effects may remain undetected.
For this situation, we propose a penalized Cox regression model with a weighted version of the Cox partial likelihood that includes patients of all subgroups but assigns them individual weights based on their subgroup affiliation. 
Patients who are likely to belong to the subgroup of interest obtain higher weights in the subgroup-specific model.
Our proposed approach is evaluated through simulations and application to real lung cancer cohorts.
Simulation results demonstrate that our model can achieve improved prediction and variable selection accuracy over standard approaches.
\end{abstract}


\section{Introduction}
\label{s:intro}

Survival analysis is an important field of biomedical research, particularly cancer research. 
The main objectives are the prediction of a patient's risk and the identification of new prognostic biomarkers to improve patients' prognosis. 
In recent years, molecular data such as gene expression data have increasingly gained importance in diagnosis and prediction of disease outcome.
Technologies for the measurement of gene expression have made rapid progress and the use of high-throughput technologies allows simultaneous measurements of genome-wide data for patients, resulting in a vast amount of data.

A typical characteristic of this kind of high-dimensional data is that the number of genomic predictors greatly exceeds the number of patients ($p >> n$). 
In this situation, the number of genes associated with a clinical outcome, here time-to-event endpoint, is typically small. 
Important objectives in modeling high-dimensional data are good prediction performance and finding a subset of predictors that are truly relevant to the outcome. 
A sparse model solution may reduce noise in estimation and increase interpretability of the results.
Another problem with high-dimensional data is that standard approaches for parameter estimation in regression models cannot handle such a large number of predictors; conventional regression techniques may not provide a unique solution to maximum likelihood problems or may result in an overfitted model.
During the last years, different approaches have been proposed for handling the $p >> n$ situation, often implying automatic variable selection, such as regularization \cite{verweij1994,tibshirani1997,zou2005} or boosting algorithms \cite{hothorn2006a,hothorn2006b,tutz2006,binder2008}. 

In clinical practice, patient cohorts are typically small.
However, when data of several patient cohorts or subgroups with the same or similar disease are available it can be reasonable to use this information and appropriately combine the data.
In multicenter studies, patients of all subgroups are often simply pooled.
When subgroups are heterogeneous with regard to their relationship between predictors and outcome, this combined analysis may suffer from biased results and averaging of subgroup-specific effects.
Standard subgroup analysis, on the other hand, includes only patients of the subgroup of interest and may lead to a loss of power when the sample size is small.

We aim at providing a separate prediction model for each subgroup that allows for identifying common as well as subgroup-specific effects and has improved prediction accuracy over both standard approaches.
Therefore, we propose a Cox proportional hazards model that allows sharing information between subgroups to increase power when this is supported by data.
We use a lasso penalty for variable selection and a weighted version of the Cox partial likelihood that includes patients of all subgroups but assigns them individual weights based on their subgroup affiliation.  
Patients who are likely to belong to the subgroup of interest obtain higher weights in the subgroup-specific model.
We estimate individual weights for each patient from the training data following the idea of \cite{bickel2008}.

We assume subgroups are known and determined by multiple cancer studies.
However, our approach can be applied to any other type of subgroups, for example, defined by clinical covariates.
Our proposed model is evaluated through simulations and application to real lung cancer cohorts, and compared to the standard subgroup model and the standard combined model.

In the following, we report on related work. In the methods Section~\ref{s:model}, we introduce the weighted Cox regression model, the estimation of patient-specific weights and the evaluation of the prediction performance. In Section~\ref{s:pipeline} the analysis pipeline for the application of our method to simulated and real data is presented. The results of the simulation study are described in Section~\ref{s:simres} and the results of the application to a set of four non-small cell lung cancer (NSCLC) cohorts in Section~\ref{s:applLC}. Section~\ref{s:discuss} contains a summary of the insights of our study and a discussion of the implications.

\subsection{Related work}

Different approaches have been published recently suggesting the use of weights in regression models to consider subgroups.
\cite{weyer2015} aim at improving stability and prediction quality of a Cox model for a specific subgroup by including one additional weighted subgroup.
The authors use a weighted and stratified Cox regression model based on componentwise boosting for automatic variable selection. They study the effects of a set of different fixed weights $w \in(0,1)$ for the additional subgroup, while all observations in the subgroup of interest obtain a weight of 1 in the stratum-/ subgroup-specific likelihood. 
In this paper, we compare a set of different fixed weights as suggested by \cite{weyer2015} to our more flexible approach with individual subgroup-specific weights estimated from the data. 
However, we assume the same baseline hazard rate across all subgroups in contrast to the stratified Cox model by \cite{weyer2015}.

Alternatively, subgroup weights can be considered as a tuning parameter and optimized by model-based optimization (MBO) to improve prediction performance in the Cox model. 
This approach by \cite{Richter2019} is also more flexible than the one by \cite{weyer2015} since it allows different fixed weights for different subgroups in each subgroup model. 
MBO helps to identify the best combination of fixed weights with regard to prediction accuracy.

Bayesian approaches for the estimation of subgroup weights were proposed by \cite{bogojeska2012} and \cite{simon2002}. 
However, they are not designed for our high-dimensional situation since they do not perform variable selection.

Weighted regression models are also used in local regression, however without predefined groups.
For each individual, a local regression model is fitted based on its neighboring observations. 
The latter are weighted by their distances from the observation of interest. 
Penalized localized regression approaches for dealing with high-dimensional data exist \cite{tutz2005, binder2012}.
Instead of using distance in covariate space, our proposed weights correspond to the relationship between covariates and subgroup membership.
A drawback of localized regression is that it does not provide global regression parameters, making interpretation difficult. 
Furthermore, only a small number of observations is used for each local fit in contrast to our approach, where the weighted likelihood is based on all training data. 

We define subgroups by multiple cancer studies or cohorts and aim at appropriately combining them to increase power and simultaneously, considering heterogeneity among the subgroups.
This idea of combining data from different data sources is similar to integrative analysis.
In high-dimensional settings with genomic predictors, different publications suggest the use of specific penalties in regularized regression for parameter estimation and variable selection across multiple data types.
For example, \cite{liu2014} and \cite{liu2014b} propose composite penalties with two-level gene selection. In the first selection level represented by an outer penalty, the association of a specific gene in at least one study is determined. In the second level, inner penalties of ridge or lasso type are used to allow the selection of either the same set of genes or different sets of genes in all studies.
Instead of aggregating multiple studies with the same type of (omics) data, \cite{boulesteix2017} perform an integrative analysis of multiple omics data types available for the same patient cohort. 
The authors use a lasso penalty with different penalty parameters for the different data types.
\cite{bergersen2011} integrate external information provided by another genomic data type by using a weighted lasso that penalizes each covariate individually with weights inversely proportional to the external information.
\cite{gade2011} use a bipartite graph to integrate miRNA and gene expression data from the same patient cohort into one prediction model to find a combined signature that improves the prediction. 
This graph is built by combining correlations between both data types and external information on target predictions.


\section{Methods}
\label{s:model}

\subsection{Cox proportional hazards model}

Assume the observed data of patient $i$ consists of the tuple $(t_i, \delta_i)$, the covariate vector $\boldsymbol{x}_i=(x_{i1},\ldots,x_{ip})' \in \mathds{R}^{p}$, and the subgroup membership $s_i \in \{1,\ldots,S\}$ with $S$ the number of subgroups in the complete data set, and $i=1,\ldots,n$. 
$t_i=\min(T_i,C_i)$ denotes the observed time of patient $i$, with $T_i$ the event time and $C_i$ the censoring time.
$\delta_i = \mathds{1}(T_i \leq C_i)$ indicates whether a patient experienced an event ($\delta_i=1$) or was (right-)censored ($\delta_i=0$).

The most popular regression model in survival analysis is the Cox proportional hazards model \cite{cox_regression_1972}.
It models the hazard rate $h(t|\boldsymbol{x}_i)$ of an individual at time $t$ as
\[
  h(t|\boldsymbol{x}_i)= h_0(t) \cdot \exp(\boldsymbol{\beta}' \boldsymbol{x}_i ) = h_0(t) \cdot \exp\left(\sum_{j=1}^p \beta_j x_{ij}\right),
\]
where $h_0(t)$ is the baseline hazard rate, and $\boldsymbol{\beta}=(\beta_1,\ldots,\beta_p)'$ is the unknown parameter vector. 
The regression coefficients $\beta_j$ are estimated by maximizing a partial likelihood without having to specify the baseline hazard rate.

\subsection{Penalized Cox regression model}

We consider high-dimensional settings where the number of covariates $p$ exceeds the sample size $n$. 
In this situation, the solution maximizing the Cox partial likelihood is not unique. 
One possibility to deal with this problem is to introduce a penalty term into the partial log-likelihood $l(\boldsymbol{\beta})$, referred to as regularization. 
This approach is also reasonable in $p < n$ settings since it considers collinearity among the predictors and helps to prevent overfitting. 
We use a lasso penalty \cite{tibshirani1996,tibshirani1997} that performs variable selection and yields a sparse model solution.
The resulting maximization problem of the penalized partial log-likelihood is given by
\[
  \hat{\boldsymbol{\beta}} = \underset{\boldsymbol{\beta}}{\text{argmax}} \left\{ l(\boldsymbol{\beta}) - \lambda \cdot \sum_{j=1}^p |\beta_j| \right\} .
\]
The parameter $\lambda$ controls the strength of penalization and is optimized by 10-fold cross-validation.
For parameter estimation, we use the implementation in the R package glmnet \cite{friedman2010}.

\subsection{Weighted Cox partial likelihood}

In the standard unweighted partial likelihood, all patients contribute to the same extent to the estimation of the regression coefficients. 
This might not be desirable when the cohort is heterogeneous due to known subgroups that are associated with different prognosis. 
In this situation, it is reasonable to fit a separate Cox model for each subgroup. 
This can be done by using only the data from the subgroup of interest or by including information from the other subgroups. 
We include patients from all subgroups in the likelihood for one specific subgroup but assign them individual weights $w_i \in [0, 1]$, $i = 1,\ldots,n$ to account for the heterogeneity in the data.
The size of each weight determines to which extent the corresponding patient contributes to the estimation.
 
In accordance with \cite{weyer2015}, the weighted version of the partial log-likelihood is defined as
\begin{align*}
  l(\boldsymbol{\beta}) = \sum_{i=1}^n \delta_i w_i \Big( \boldsymbol{\beta}' \boldsymbol{x}_i  - 
   \ln \Big[ \sum_{k=1}^n \mathds{1}(t_i \leq t_k) w_k \exp \left( \boldsymbol{\beta}' \boldsymbol{x}_k \right) \Big] \Big) .
\end{align*}
\cite{weyer2015} propose the use of fixed weights.
The idea is to focus on a specific subgroup $s$ of patients and assign each of these patients a weight of 1, while all other patients are down-weighted with a fixed weight $w \in (0,1)$:
\[
w_i = \begin{cases}
1, & \text{if } s_i = s \\
w, & \text{else}.
\end{cases}
\]
Standard subgroup analysis is based only on the patients in the subgroup of interest $s$, which corresponds to $w=0$ for all patients not belonging to $s$.
A combined model that pools patients from all subgroups corresponds to $w=1$ for all patients.
Alternatively to the idea of \cite{weyer2015}, we propose to estimate individual weights for each patient from the training data.
This approach is described in the following Section \ref{s:estwei}.

\subsection{Estimation of weights}
\label{s:estwei}

Individual weights for each patient in each subgroup-specific likelihood can be estimated from the training data following the idea of \cite{bickel2008}.
The weights match the joint distribution of all subgroups to the target distribution of a specific subgroup $s$,
such that a patient who is likely to belong to the subgroup of interest receives a higher weight in the subgroup-specific model.

Assume the entire training data from all subgroups are summarized in the covariates $\boldsymbol{x}$ and a response $\boldsymbol{y}$. 
In time-to-event settings, the response $y_i$ corresponds to the tuple $(t_i,\delta_i)$, with $t_i$ the observed time until an event or censoring and $\delta_i$ the event indicator. 
Let $\ell(\boldsymbol{y}, f_s(\boldsymbol{x}))$ be an arbitrary loss function and $f_s(\boldsymbol{x})$ the predicted response based on the observed covariates in subgroup $s$.
$f_s(\boldsymbol{x})$ should correctly predict the true response and thus minimize the expected loss with respect to the unknown joint distribution $p(\boldsymbol{y},\boldsymbol{x}|s)$ for each subgroup $s$, given by $\text{E}_{p(\boldsymbol{y},\boldsymbol{x}|s)}[ \ell(\boldsymbol{y},f_s(\boldsymbol{x}))]$.
The following equation shows that this expected loss for each subgroup equals the expected weighted loss with respect to the joint distribution of the pooled data from all subgroups $p(\boldsymbol{y},\boldsymbol{x})$
\begin{align*}
\text{E}_{p(\boldsymbol{y},\boldsymbol{x}|s)}[ \ell(\boldsymbol{y},f_s(\boldsymbol{x}))] &= \int p(\boldsymbol{y},\boldsymbol{x}|s) \ell(\boldsymbol{y},f_s(\boldsymbol{x})) \text{d}\boldsymbol{y} \text{d}\boldsymbol{x}  \\
&= \int \frac{p(\boldsymbol{y},\boldsymbol{x}|s)}{p(\boldsymbol{y},\boldsymbol{x})}p(\boldsymbol{y},\boldsymbol{x}) \ell(\boldsymbol{y},f_s(\boldsymbol{x})) \text{d}\boldsymbol{y} \text{d}\boldsymbol{x}  \\
&= \text{E}_{p(\boldsymbol{y},\boldsymbol{x})} \left[ \frac{p(\boldsymbol{y},\boldsymbol{x}|s)}{p(\boldsymbol{y},\boldsymbol{x})} \ell(\boldsymbol{y},f_s(\boldsymbol{x})) \right]  \\
&= \text{E}_{p(\boldsymbol{y},\boldsymbol{x})} \left[ w_s(\boldsymbol{y},\boldsymbol{x}) \ell(\boldsymbol{y},f_s(\boldsymbol{x})) \right].
\end{align*}
The subgroup-specific weights for each patient are defined as 
\[
w_s(\boldsymbol{y},\boldsymbol{x}) = \frac{p(\boldsymbol{y},\boldsymbol{x}|s)}{p(\boldsymbol{y},\boldsymbol{x})} = \frac{p(s|\boldsymbol{y},\boldsymbol{x})}{p(s)}  , \quad p(s)>0  .
\]
The last equation shows that the weights can be expressed in terms of $p(s)$ and $p(s|\boldsymbol{y},\boldsymbol{x})$.
$p(s)$ can be estimated by the relative frequency of subgroup $s$ in the overall training cohort, and
$p(s|\boldsymbol{y},\boldsymbol{x})$ can be considered as a multi-class classification problem \cite{bickel2008}. 
We estimate $p(s|\boldsymbol{y},\boldsymbol{x})$ by multinomial logistic regression or by random forest, using the implementation in the R packages glmnet \cite{friedman2010} and ranger \cite{wright2017}, respectively. 
Unlike \cite{bickel2008}, we use 10-fold cross-validation to estimate $p(s|\boldsymbol{y},\boldsymbol{x})$ from the training data to prevent overfitting.
As a result, for each subgroup, we obtain an $n$-dimensional vector of estimated weight for each patient.

\subsection{Prediction performance}

Prediction performance of all Cox models is evaluated by Harrel's C- (concordance) index \cite{harrell1996}, implemented in the R package Hmisc \cite{harrell2018}. 
The C-index is a measure of predictive discrimination and defined as the proportion of all usable pairs of patients with concordant predicted and observed survival times. 
Let $t_i$, $t_{i^\ast}$ be the observed survival times of patients $i$ and $i^\ast$, and ${\hat{r}(\boldsymbol{x}_i)=\boldsymbol{\hat{\beta}}'\boldsymbol{x}_i}$, ${\hat{r}(\boldsymbol{x}_{i^\ast})=\boldsymbol{\hat{\beta}}'\boldsymbol{x}_{i^\ast}}$ the corresponding risk scores.
A pair $(i, i^\ast)$ is considered concordant if \, ${t_i \lesseqgtr t_{i^\ast} \Leftrightarrow \hat{r}(\boldsymbol{x}_i) \gtreqless \hat{r}(\boldsymbol{x}_{i^\ast})}$.
The C-index is defined as
\[
\textsl{CI} = \frac{1}{n_c} \sum_{ \{i: \, \delta_i=1\} } \sum_{ \{i^\ast: \, t_{i^\ast} > t_i \} } \left( \mathds{1}(\hat{r}(\boldsymbol{x}_{i^\ast})<\hat{r}(\boldsymbol{x}_i)) + \frac{1}{2} \mathds{1}(\hat{r}(\boldsymbol{x}_{i^\ast})=\hat{r}(\boldsymbol{x}_i)) \right) ,
\]
where $n_c$ is the number of comparable pairs $(i, i^\ast)$ that standardizes \textsl{CI} to $[0,1]$.
A patient pair is considered unusable, if both patients die at the same time, or both patients are censored, or if one is censored before the other one dies. 
$\textsl{CI} \approx 1$ stands for a very good prediction and values around 0.5 suggest a random prediction.

In previous simulation studies and the same application example we also considered the Brier Score as a measure of prediction accuracy. We found that both measures, Brier Score and C-index led to similar results and the same conclusions \cite{madjar2018}.


\section{Model fitting and evaluation}
\label{s:pipeline}

We compare our weighted approach with the standard (unweighted) models, i.e.\ the combined model and the subgroup model, as well as a weighted Cox model with fixed weights as proposed by \cite{weyer2015}.
In the latter, patients belonging to a certain subgroup are assigned a weight of 1 in the subgroup-specific likelihood, while all other observations are down-weighted with a constant weight $w \in (0,1)$. 
For our proposed approach we compare three different classification methods for weights estimation with respect to prediction performance: Multinomial logistic regression with lasso (\textit{lasso}) or ridge (\textit{ridge}) penalty, and random forest (\textit{rf}).
All Cox models include a lasso penalty for variable selection. 
We compare the following Cox models concerning prediction performance. 
The italic expressions in parentheses denote the abbreviations of the models in the following analyses: 
\begin{itemize}
\item Weighted model with estimated weights (\textit{lasso}, \textit{ridge}, \textit{rf})
\item Weighted model with fixed weights ($w=0.1,0.2,...,0.9$)
\item Standard subgroup model (\textit{sub}), using only patients of a specific subgroup
\item Standard combined model (\textit{all}), using patients of all subgroups. The subgroup indicator is included as additional covariate.
\end{itemize}

\begin{figure}[!htb]
	\centering
  \includegraphics[width=.8\textwidth]{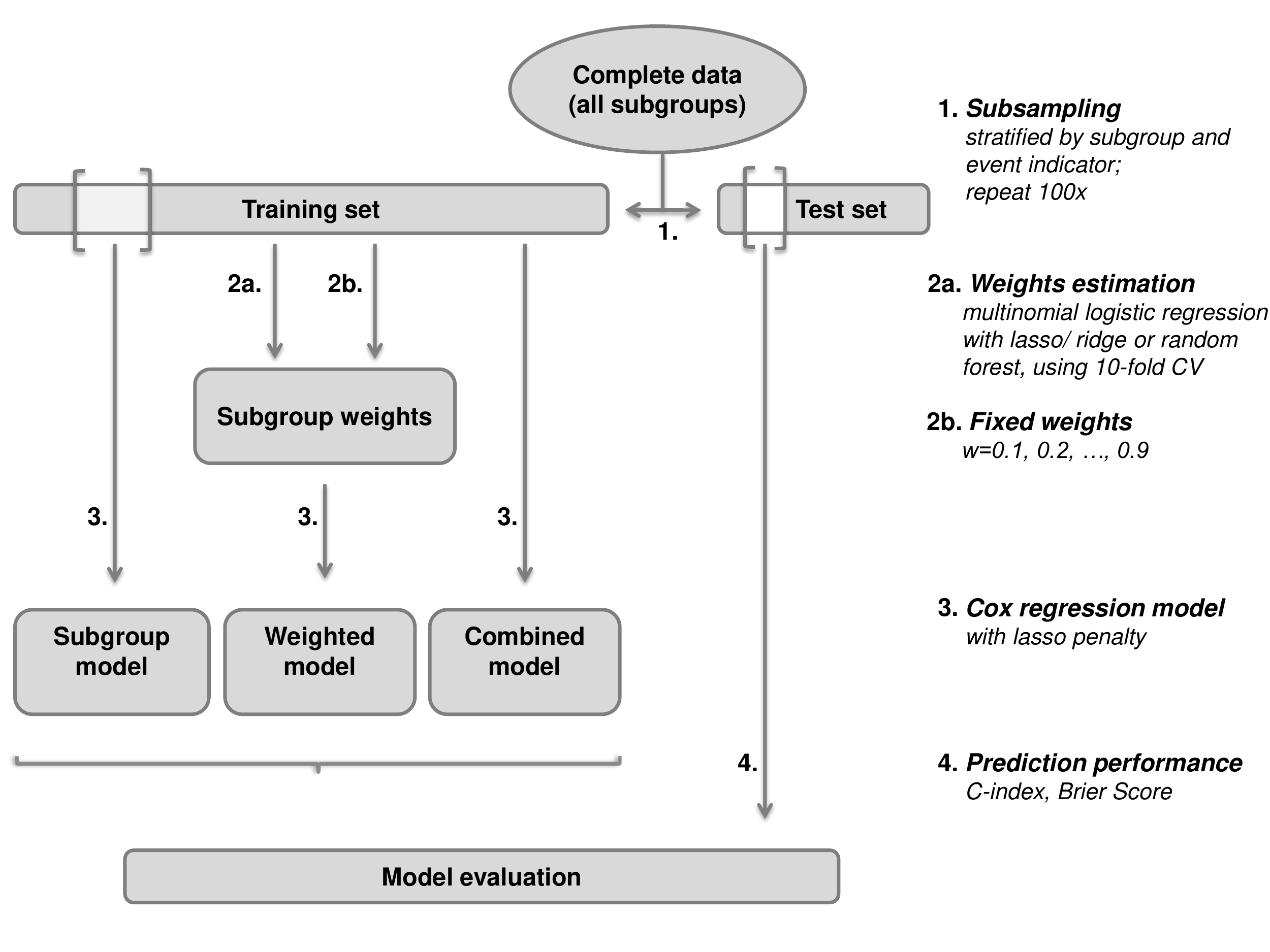} 
\caption{Analysis pipeline for the simulation study; Brighter regions in the training and test set indicate the observations of the subgroup.}
\label{fig_pipeline}
\end{figure}

Figure \ref{fig_pipeline} provides a schematic representation of the analysis pipeline.
First, we randomly generate training data sets for model fitting and test data sets for model evaluation and repeat this procedure 100 times.
In the application example, we repeatedly randomly split the complete data into training (with proportion 0.632) and test sets. 
We perform subsampling stratified by subgroup and event indicator, to take different subgroup sizes and censoring proportions into account.
In the simulation study, we repeatedly randomly generate training and test sets of the same size and with the same simulation parameters.
Second, we estimate individual subgroup weights from the training data using different classification methods and 10-fold cross-validation (CV). 
Next, we fit the combined and weighted Cox models based on the training data of all subgroups, while the standard subgroup model is based on the training data of the respective subgroup only. 
Finally, we evaluate the prediction performance of the estimated Cox models with respect to a certain subgroup using only the test data of this particular subgroup. 
The R package batchtools \cite{lang2017} is used for parallelization and the R package mlr \cite{bischl2016} is used as a framework for weights estimation, Cox model fitting and evaluation by the C-index.

\section{Simulation results}
\label{s:simres}

\subsection{Simulated data}

We simulate four subgroups (1A, 1B, 2A, 2B) of equal size $n$ from two differently distributed groups denoted by the index $g^\ast=1,2$: group~1 including subgroups 1A and 1B, and group~2 including subgroups 2A and 2B. 
Within each group we use the same parameters for the simulation of the data. 
We simulate the survival data from a Weibull distribution according to \cite{bender2005}, with scale parameter $\eta_{g^\ast}$ and shape parameter $\kappa_{g^\ast}$ estimated from two independent lung cancer cohorts (GSE37745 and GSE50081). 
For this purpose, we compute survival probabilities at 3 and 5 years using the Kaplan-Meier estimator for both lung cohorts separately.
The corresponding probabilities are 57\% and 75\% for 3-years survival, and 42\% and 62\% for 5-years survival, respectively.
Individual event times in group $g^\ast$ are simulated as
\[
	T_{g^\ast} \sim \left(- \frac{\log(U)}{\eta_{g^\ast} \exp(\boldsymbol{x}_{g^\ast} \boldsymbol{\beta}_{g^\ast})}\right)^{1/\kappa_{g^\ast}} , \quad U \sim \mathcal{U}[0,1],
\]
with true effects $\boldsymbol{\beta}_{g^\ast} \in \mathds{R}^{p}$, $g^\ast=1,2$. 
We randomly draw noninformative censoring times $C_{g^\ast}$ from a Weibull distribution with the same parameters as for the event times, resulting in approximately 50\% censoring rates in both groups.
The individual observed event indicators and times until an event or censoring are defined as $\delta_{g^\ast} = \mathds{1}(T_{g^\ast} \leq C_{g^\ast})$ and  $t_{g^\ast}=\min(T_{g^\ast},C_{g^\ast})$.

For each subgroup we simulate $p$ uncorrelated (genetic) covariates $\boldsymbol{x}_{g^\ast}$ from a multivariate normal distribution with mean vector $\boldsymbol{\mu}_{g^\ast}$ and covariance matrix $\boldsymbol{\Sigma}=\boldsymbol{I}_{p\times p}$. 
In previous simulation studies we compared the results of different covariance structures and found no remarkable differences \cite{madjar2018}.
Elements of $\boldsymbol{\mu}_{g^\ast}$ are defined by a linear function with parameter $\epsilon \in [0,1]$ that reflects the degree of similarity between the two groups. 
We assign $\mu = 4+4\cdot\epsilon$ to genes with a strong effect on the outcome ($|\beta|=1$), $\mu = 4+2\cdot\epsilon$ corresponds to genes with a moderate effect ($|\beta|=0.5,0.75$), and $\mu = 4$ to genes with a weak or no effect ($|\beta|=0,0.25$).
This choice relies on the assumption that prognostic genes have a higher expression level than noise genes.
The magnitude of $\mu$ is chosen following real gene expression data, where the expression values typically range from 4 to 12 after transformation to $\log_2$ scale.

In all simulated scenarios, we assume the first 12 genes to be prognostic in at least one of the two groups. The true effects are given by
	\begin{center}
		\begin{tabular}{c|cccccccccccc}
 	Gene & 1 & 2 & 3 & 4 & 5 & 6 & 7 & 8 & 9 & 10 & 11 & 12 \\
 \hline
$\boldsymbol{\beta}_1$ & 1 & 1 & 0 & 0 & -0.5 & 0.5 & 0.75 & 0.25 & -1 & -1 & -0.75 & -0.25  \\
$\boldsymbol{\beta}_2$ & 0 & 0 & 1 & 1 & 0.5 & -0.5 & 0.25 & 0.75 & -1 & -1 & -0.75 & -0.25  \\
\end{tabular} 
	\end{center}
We include subgroup-specific effects (genes 1 to 4), opposite effects (genes 5 and 6), effects in the same direction but of different size (genes 7 and 8), and joint effects of varying sizes (genes 9 to 12).
We choose these effects with alternate signs so that they sum up to zero, resulting in reasonable simulated survival times.
In settings with $p>12$, we assume all remaining genes to represent noise and being unrelated to the survival times in both groups (${\beta_{13}=...=\beta_p=0}$). 

In our simulation study we focus on high-dimensional settings where the sample size $n$ is small compared to the number of covariates (genes) $p$, a typical characteristic of gene expression data.
We consider the following parameters for the data simulation, resulting in 252 different combinations in total: 
\begin{center}
\begin{tabular}{l|l}
Parameter & Values (per subgroup) \\
\hline
$n$ & 20, 30, 40, 50, 60, 70, 80, 90, 100, 200, 500, 1000 \\  
$p$ & 12, 100, 200 \\ 
$\epsilon$ & 0, 0.1, 0.2, 0.3, 0.4, 0.5, 1 \\
\end{tabular} 
	\end{center}

\subsection{Weights estimation}

Our proposed subgroup model uses patients from all subgroups for training but assigns them individual weights in the Cox partial likelihood based on their subgroup membership.
Weights for a specific subgroup are estimated by the individual predicted probabilities of belonging to this subgroup, obtained by classification, divided by the subgroup proportion.
Thus, a patient who is likely to belong to the subgroup of interest receives a higher weight in the subgroup-specific likelihood. 
We compare three different classification methods that are appropriate for multi-class problems and high-dimensional covariates with respect to their predictive quality and their ability to discriminate between differing subgroups. 

\begin{figure}[htb]
	\centering
  \includegraphics[page=1, width=.45\textwidth]{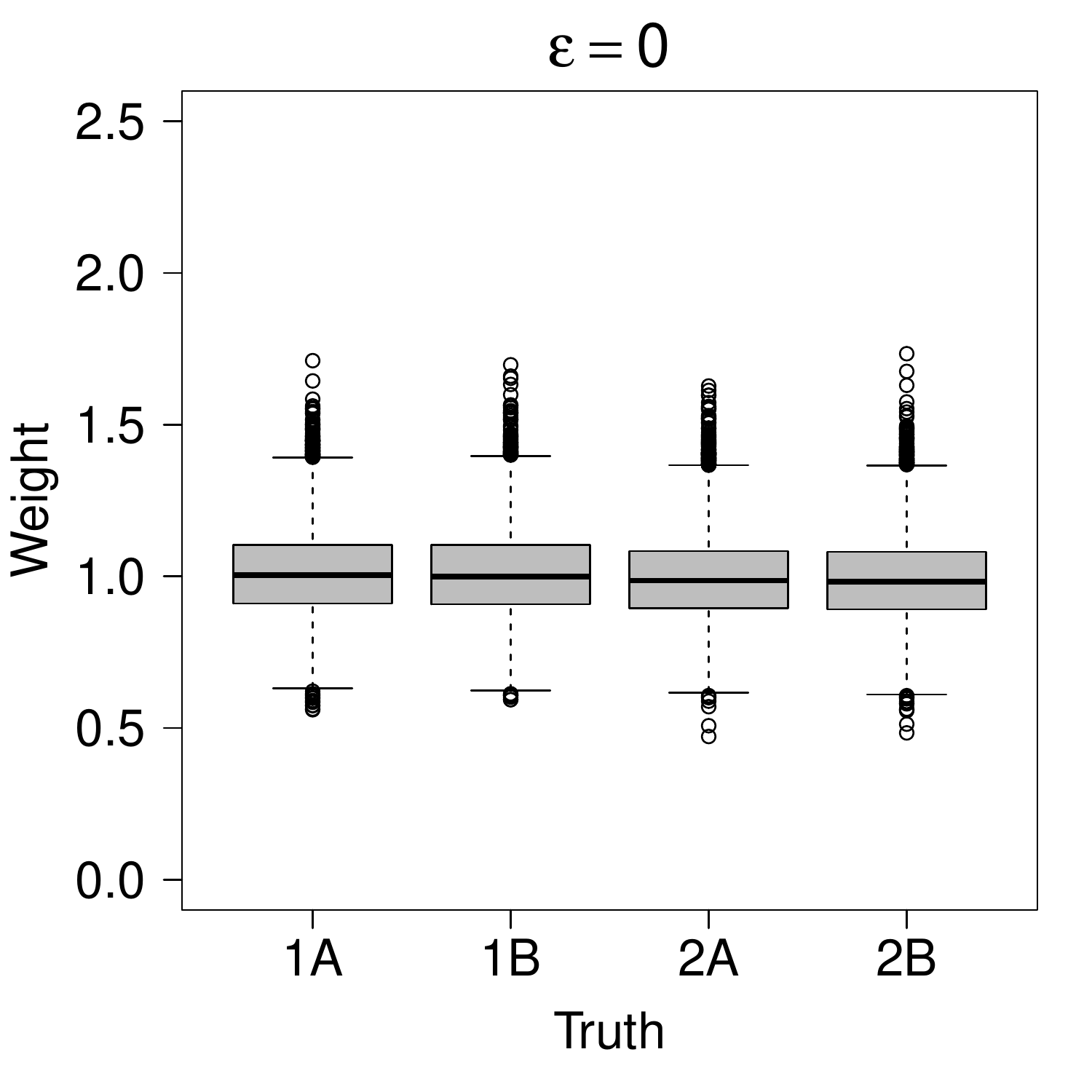} 
  \includegraphics[page=6, width=.45\textwidth]{Boxplots_weights_p100_n100_rF} 
	\caption{Estimated weights for subgroup 1A obtained by random forest based on simulated training data with $p=n=100$ and $\epsilon=0,0.5$.}
\label{fig_sim_wei}
\end{figure}

Figure \ref{fig_sim_wei} displays boxplots of the estimated weights for subgroup 1A across all training sets in two selected simulation scenarios with $\epsilon=0$ and $\epsilon=0.5$. 
The $x$-axis represents the true subgroup membership of each observation, and the $y$-axis the individual weights estimated by random forest (\textit{rf}) for subgroup 1A. 
Results of all three classification methods (\textit{lasso}, \textit{ridge}, \textit{rf}) are relatively similar, altough \textit{rf} tends to perform best for small $\epsilon$ and $n$, whereas for large sample size the discriminative ability of \textit{lasso} and \textit{ridge} is slightly better. 
The largest difference in results is obtained for different values of $\epsilon$.
When all subgroups are very similar ($\epsilon=0$), multi-class classification fails to distinguish the two differing groups. 
All observations are assigned a weight of approximately~1 in all subgroup models, similar to the standard combined Cox model.
The corresponding area under the ROC curve (AUC) for the distinction between group 1 and 2 (computed based on test data and cross-validated training data) is approximately 0.5, indicating that prediction performance is not much better than random (see Figure~\ref{fig_sim_auctest}). 
Increasing values of $\epsilon$, meaning larger differences between the two groups, lead to improved prediction performance (see Figure~\ref{fig_sim_auctest}), and for $\epsilon=0.5$ classification succeeds in providing an almost perfect separation between both groups with $\text{AUC} \approx 1$.
Larger sample size $n$ and smaller number of covariates $p$ also result in better prediction performance.

\subsection{Parameter estimation and prediction performance}

Weighted Cox models, including fixed or estimated weights (with different classification methods for weights estimation), are compared
to the standard combined and subgroup model, first by estimated regression coefficients and second by prediction performance.

\begin{figure}[!htb]
	\centering
  \includegraphics[page=1, width=.8\textwidth]{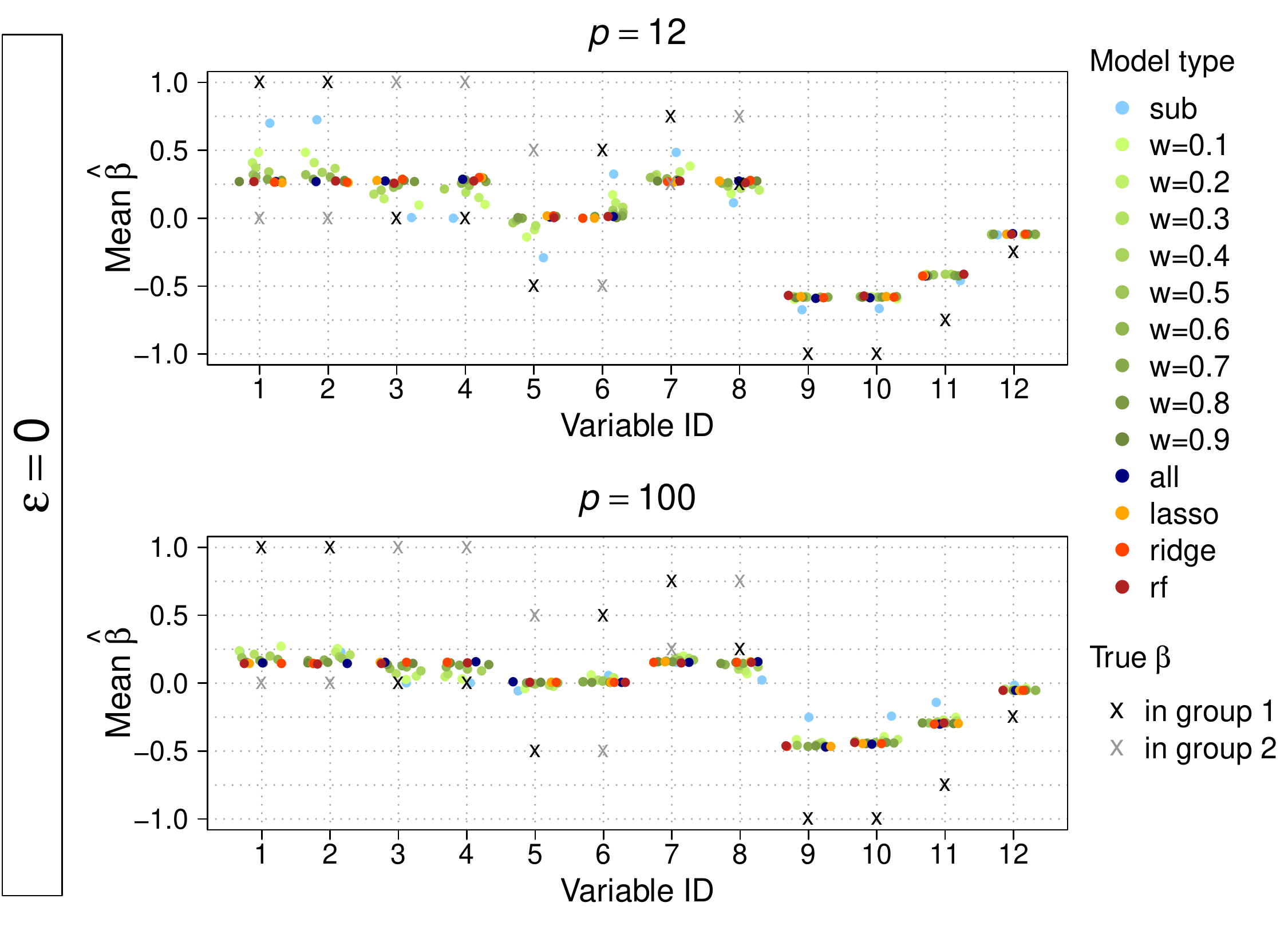} \\
	\includegraphics[page=2, width=.8\textwidth]{Scatterplot_Betas_CoxM_N50-2} 
	\caption{Mean estimated regression coefficients of the first 12 prognostic genes in all Cox models for group 1 (averaged across all simulated training data sets, and subgroups 1A and 1B) for $n=50$ and $\epsilon=0,0.5$.} 
\label{fig_sim_betas}
\end{figure}

Figure \ref{fig_sim_betas} shows scatterplots of the mean estimated regression coefficients of the first 12 prognostic genes in group 1 (mean across all training sets
and subgroups 1A and 1B) for simulated data with $n=50$, $p=12,100$ and $\epsilon=0,0.5$.
For $\epsilon=0$, the combined and weighted model with estimated weights provide very similar results, as expected. 
They identify joint effects better than the subgroup model when the sample size is small ($n \leq p$) and otherwise equally well. 
However, the subgroup model estimates subgroup-specific effects better, especially for increasing sample size, whereas the other two model approaches tend to average effects across all subgroups.
For larger values of $\epsilon$ the estimated weights model detects subgroup-specific effects increasingly better than the combined model, and similarly well or even better than the standard subgroup model when sample size is small. 
Results for fixed weights lie between the subgroup model and the combined model.

These findings agree with the corresponding mean inclusion frequencies (MIFs), defined as the proportion of training data sets in which a specific covariate $j$ is included in the model ($\hat{\beta}_j \neq 0$).
For small sample size, the MIFs of the standard combined model and the estimated weights approach are larger than the MIFs of the standard subgroup model.
This has a positive impact on the detection of joint effects, but subgroup-specific effects that are present in only one group may be more often erroneously selected in the other group. 
For increasing sample size the MIFs of all models also increase.
For larger values of $\epsilon$, the MIFs of the estimated weights model move closer to the MIFs of the subgroup model regarding subgroup-specific effects and are still similar to the combined model for joint effects. 

\begin{figure}[!htb]
	\centering
  \includegraphics[page=1, width=.8\textwidth]{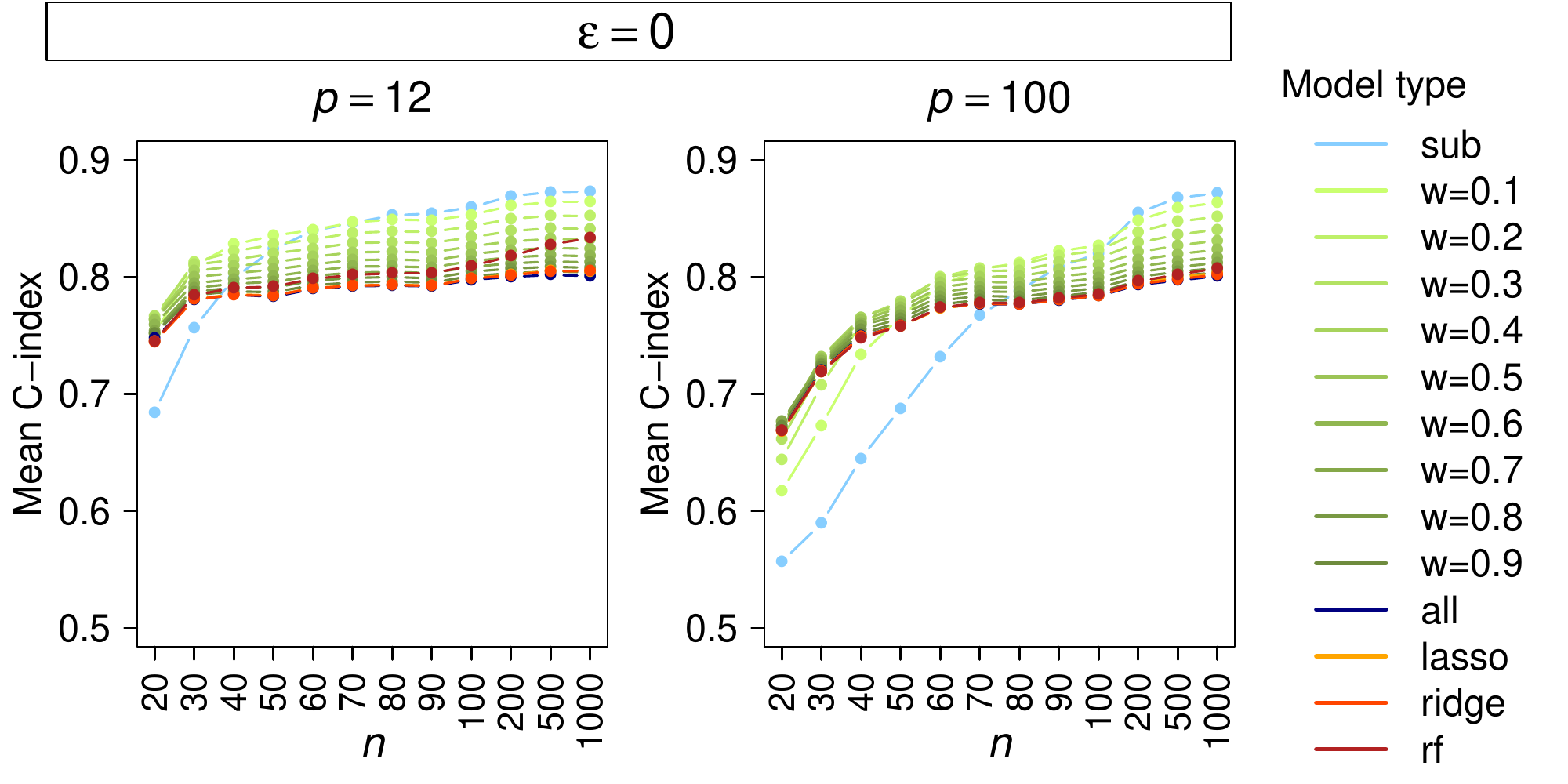} \\
	\includegraphics[page=2, width=.8\textwidth]{Mean_CIndex3} 
	\caption{Mean C-index, averaged across all test data sets and subgroups, for $\epsilon=0,0.5$.} 
\label{fig_sim_CI}
\end{figure}

Finally, we assess the prediction performance of all Cox models in terms of the C-index. 
High values of the C-index (close to 1) indicate a good predictive performance, whereas 0.5 corresponds to random prediction.
Figure \ref{fig_sim_CI} displays the mean C-index (averaged across all test sets and subgroups).
For $\epsilon=0$ the combined model and the weighted model with estimated weights exhibit a very similar predictive ability, that is better compared to the subgroup model when sample size is small.
However, when the sample size increases the subgroup model outperforms the other methods. 
For larger values of $\epsilon$, the estimated weights approach performs best when the sample size is small and otherwise equally well as the subgroup model. 
Estimated weights by \textit{lasso} and \textit{ridge} improve in comparison to \textit{rf} (random forest) for larger $n$. 
Unsurprisingly, the prediction performance of fixed weights lies between the standard combined model and the subgroup model.
Mean C-index values for all 252 simulation scenarios and all 14 Cox model types can be found in  Table~\ref{tab_sim_CI}.

\section{Application to NSCLC cohorts}
\label{s:applLC}

We apply all methods presented in Section~\ref{s:simres} to the following four non-small cell lung cancer (NSCLC) cohorts comprising in total $n=635$ patients with available overall survival endpoint and Affymetrix microarray gene expression data: GSE29013 ($n=55$, 18 events), GSE31210 ($n=226$, 35 events), GSE37745 ($n=194$, 143 events), and GSE50081 ($n=160$, 65 events). 
For the analysis, we use the total number of $p=54675$ genetic covariates measured in each cohort, as well as two preselected reduced gene sets.
One gene filter is defined by the $p=1000$ features with the highest variance in gene expression values across all four cohorts, referred to as top-1000-variance genes.
The second gene filter is a literature-based selection of $p=3429$ prognostic genes.
More details on the data description and preprocessing can be found in the Supplementary Materials.

\subsection{Weights estimation}

\begin{figure}[htb]
	\centering
  \includegraphics[page=1, scale=0.65]{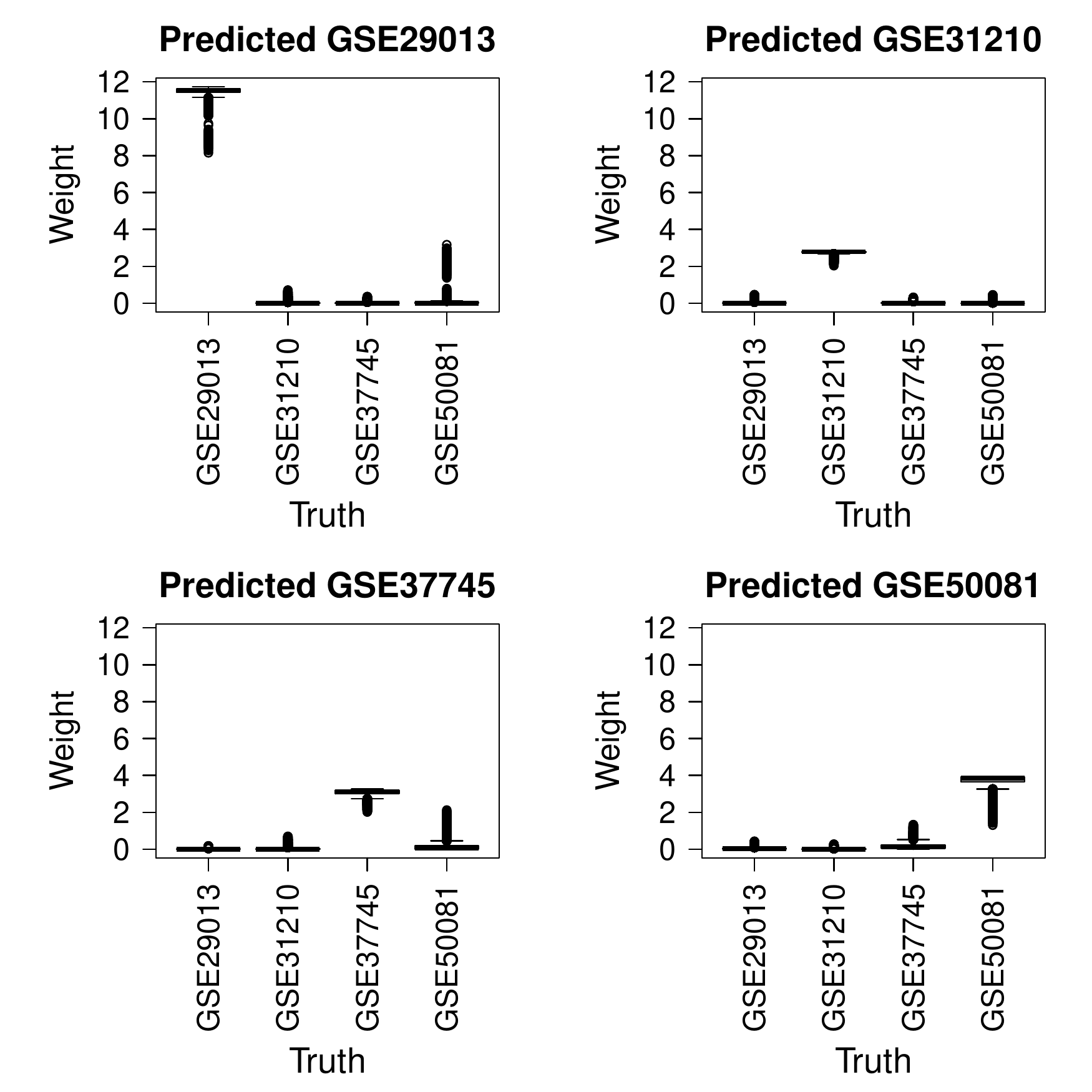} 
	\caption{Estimated weights for all lung cancer cohorts using random forest and the top-1000-variance genes as gene filter.} 
	\label{fig_LC_wei}
\end{figure}

In the following, we consider four lung cancer cohorts as subgroups.
We compare the estimated weights using three classification methods (\textit{lasso}, \textit{ridge}, \textit{rf}) and three different pre-specified sets of genes (gene filters): all available genes ($p = 54675$), top-1000-variance genes ($p = 1000$), and a literature-based selection of prognostic genes ($p = 3429$).
Since all results are very similar, we only show them exemplary for the top-1000-variance genes and \textit{rf} in Figure \ref{fig_LC_wei}.
Boxplots of the estimated weights suggest that subgroups are very different from each other.
Patients belonging to the subgroup of interest receive a relatively large weight in the respective subgroup-specific model, while the contribution of all other
subgroups is close to zero. 
This is similar to the standard subgroup model. 

\subsection{Parameter estimation and prediction performance}

All analyses are based on probe set level of gene expression data, but for the illustration of the parameter estimates in the Cox models, probe set IDs are translated into gene symbols using the R/Bioconductor annotation package hgu133plus2.db \cite{carlson2016}. 
In case of missing gene symbols, original probe set IDs are retained. 
Corresponding gene annotation is retrieved from the Ensembl website \cite{zerbino2018} to obtain gene-specific information on encoded proteins, related pathways, Gene Ontology (GO) annotations, associated diseases, and related articles in PubMed. 
This information is retrieved from the NCBI Gene \cite{brown2015} and GeneCards \cite{GeneCards2018} databases.

\begin{figure}[!htb] 
\centering
\includegraphics[page=7, width=.9\textwidth]{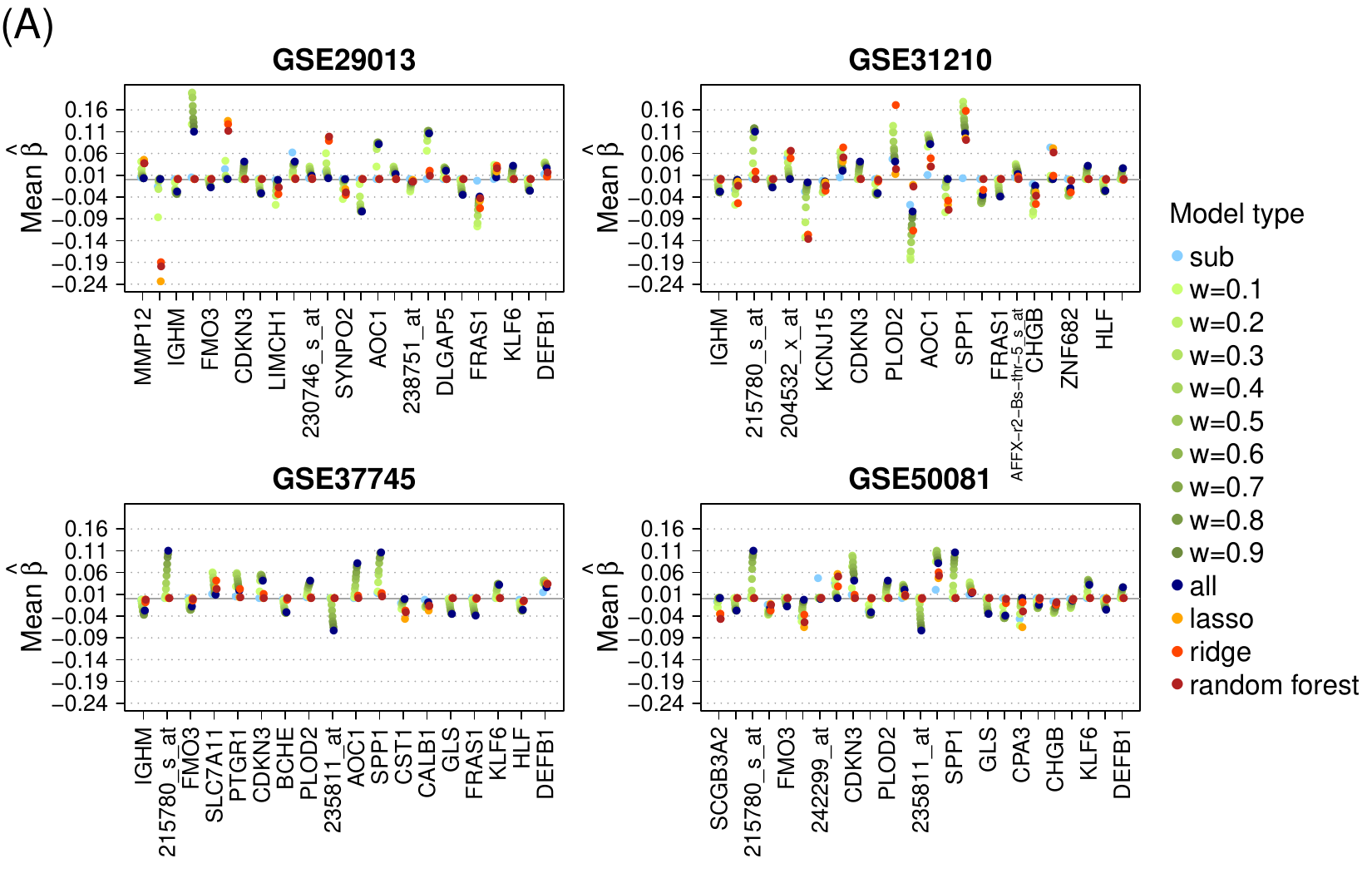}
\\
\includegraphics[page=7, width=.9\textwidth]{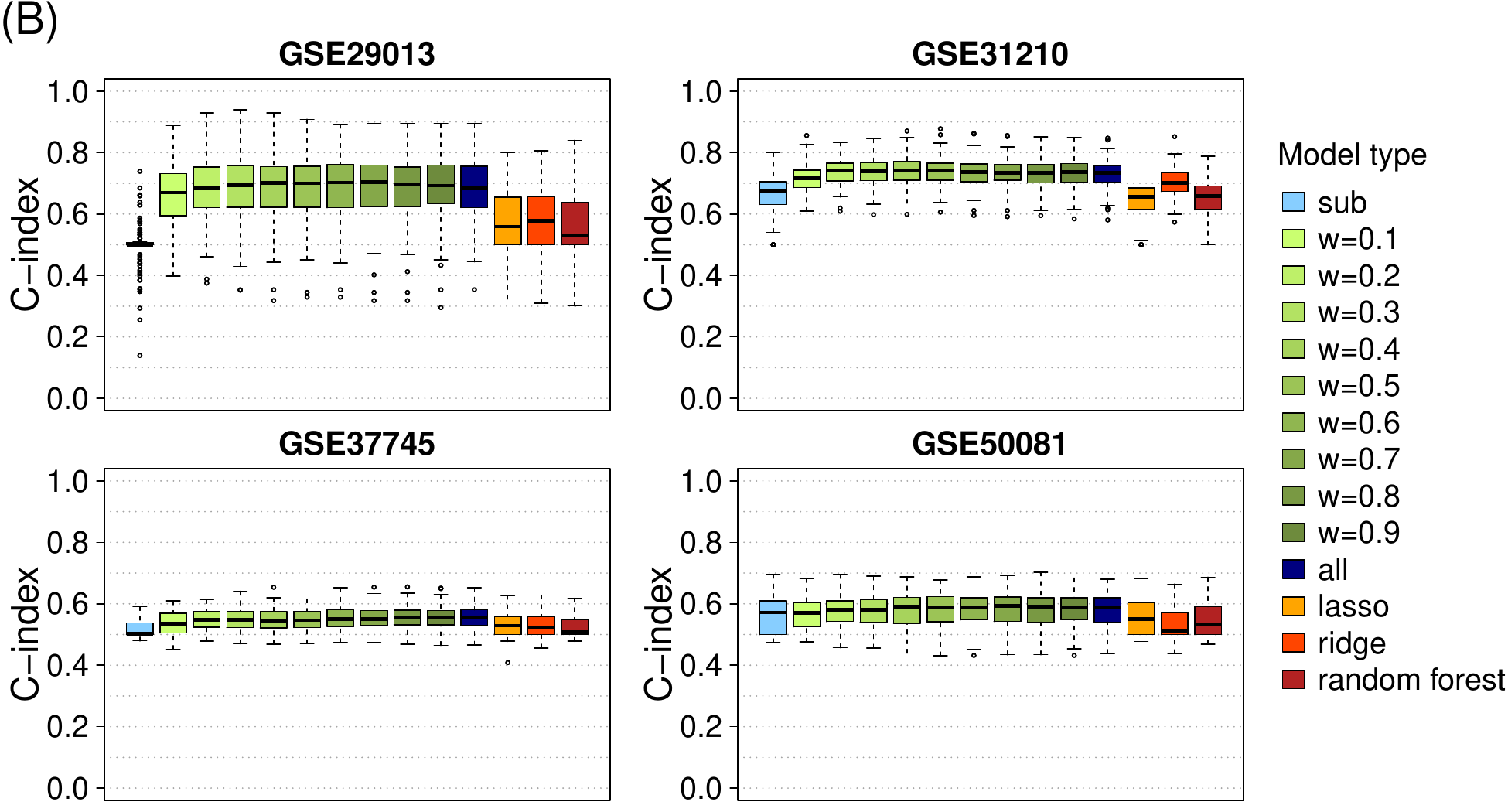}
\caption{Different types of Cox models including the top-1000-variance genes as covariates. (A) Mean estimated regression coefficients (averaged across all training sets) of selected genes. For each subgroup genes with a mean inclusion frequency larger than 0.4 in any model type are selected. (B) Boxplots of C-index based on all test sets for the prediction of each subgroup.}
\label{fig_LC_CI}
\end{figure}

Figure \ref{fig_LC_CI} shows, separately for each subgroup, the mean estimated regression coefficients of the most frequently selected top-1000-variance genes (genes with a mean inclusion frequency (MIF) larger than 0.4 in any model type).
14 genes are in the overlap of all subgroups, among them an immune-related gene (DEFB1) as well as genes (CDKN3, 215780\_s\_at/SET, GLS, KLF6, PLOD2) that were reported in the literature to be associated with different types of cancer. 
Often they are most frequently selected by the combined model and the weighted model with large fixed weights. 
Their effect estimates are positive, except for GLS.
However, the corresponding estimated regression coefficients are relatively small suggesting weak effects on survival outcome compared to the other genes included in the multivariate models. 
Subgroup-specific genes with strong effects on overall survival and high MIFs in the proposed weighted model involve the following cancer-related genes: ADH1C and BMP5 in GSE31210, as well as AREG and COL4A3 in GSE29013. 

For the other two gene filters (prognostic genes and all genes), parameter estimates of the most stable genes in all Cox models are displayed in Figures~\ref{fig_LC_betas2} and \ref{fig_LC_betas3}.
Cox models including all genes identify fewer genes compared to the other gene filters which is likely caused by the large number of noise genes. 
There are two cancer-related genes most frequently selected across all subgroups by the combined model and the weighted model with large fixed weights: ERN1 and MAGEH1. 
MIFs and estimated regression coefficients of the subgroup model and the proposed weighted model are mainly close to zero, except for PTGER3 in GSE31210. 
PTGER3 induces tumor progression in different cancer types including adenocarcinoma of the lung. This may explain the specific association with GSE31210 being the only subgroup comprising exclusively adenocarcinoma.

Interestingly, almost all selected genes are either in the overlap of all subgroups or specific for only one subgroup. There are hardly any genes selected by two or three subgroups, which may be due to the fact that these lung cancer studies are heterogeneous (see Figure~\ref{fig_LC_venn}).
There is one gene (SPP1) that is in the overlap of all four subgroups and all three gene sets. SPP1 - also known as Osteopontin (OPN) - is involved in inflammatory response, osteoblast differentiation for bone formation and attachment of osteoclasts to the mineralized bone matrix for bone resorption. Further, SPP1 is associated with several malignant diseases and prognosis in NSCLC.

Finally, all Cox models are compared with regard to prediction performance.
In Figure \ref{fig_LC_CI} results of the C-index across all test sets are shown for the top-1000-variance genes.
The combined model and fixed weights of increasing size mostly have the highest predictive accuracy, while the estimated weights approach and the standard subgroup model perform similarly badly. 
The corresponding boxplots of the C-index for the prognostic gene filter and all genes are shown in Figures \ref{fig_LC_CI2} and \ref{fig_LC_CI3}.
Random forest tends to be the best classification method in combination with prognostic genes and all genes, whereas ridge tends to perform slightly better than the other classification methods along with top-1000-variance genes.
However, overall prediction performance is mostly moderate and not much better than random.


\section{Discussion} 
\label{s:discuss}

We have focused on three major objectives: prediction of a patient’s survival, selection of important covariates, and consideration of heterogeneity in data due to known subgroups of patients. 
Specifically, we have aimed at providing a separate prediction model for each subgroup that allows the identification of common as well as
subgroup-specific effects and has improved prediction accuracy over standard approaches.
As standard approaches, we consider standard subgroup analysis, including only patients of the subgroup of interest, and standard combined analysis that pools patients of all subgroups. 
We have proposed a Cox model with lasso penalty for variable selection and a weighted version of the partial likelihood that includes patients of all subgroups but with individual weights.  
This allows sharing information between subgroups to increase power when this is supported by data, meaning that subgroups are similar in their covariates and survival outcome.
Weights for a specific subgroup are estimated from the training data by classification and cross-validation such that they represent the probability of belonging to that subgroup given the observed covariates and survival outcome. 
These predicted conditional probabilities are divided by the a priori probability of the respective subgroup to obtain the subgroup-specific weights for each patient. 
Patients who fit well into the subgroup of interest receive higher weights in the subgroup-specific model.

We have considered three different classification methods for weights estimation (multinomial logistic regression with lasso or ridge penalty and random forest), and, based on simulated data and on real data, we have compared our proposed weighted Cox model to both standard Cox models (combined and subgroup), as well as a weighted Cox model with different fixed weights as proposed by \cite{weyer2015}. 
Observations belonging to a certain subgroup were assigned a weight of 1 in the subgroup-specific likelihood, while all other observations were down-weighted with a constant weight $w \in \{0.1,0.2,...,0.9\}$.

Simulation results have shown that when subgroups were hardly distinguishable from each other with respect to their covariate values and differed mainly in their relationship between prognostic covariates and survival outcome, classification methods failed to discriminate between distinct subgroups and all observations were assigned a weight around one corresponding to the standard combined model.
In these situations, results of the combined model and the proposed weighted model were very similar.
Both models had better prediction performance and larger power to detect joint effects than the standard subgroup model when the sample size was small ($n\leq p$).
However, they tended to average subgroup-specific effects across subgroups, leading to biased estimates.

For increasing sample size, the standard subgroup model outperformed the other models regarding prediction and selection accuracy, in particular in terms of correct estimation of subgroup-specific effects.
When differences between subgroups became larger, classification succeeded in discriminating between different subgroups, and our proposed weighted model improved over the combined model in correctly identifying subgroup-specific effects and resulted in higher prediction accuracy. 
It clearly outperformed the standard subgroup model when the sample size was low, and otherwise performed similarly well. 
Results with fixed weights, as expected, always lay between the standard subgroup model and the combined model. 

In the application example, we considered four lung cancer studies as subgroups comprising overall survival outcome, and gene expression data as covariates.
Three different gene filters were used: all available genes, top-1000-variance genes, and a literature-based selection of prognostic genes. 
Estimated weights suggested large differences between all subgroups and resembled the standard subgroup model, where only the
subgroup of interest is assigned a high weight and all other subgroups have weights close to zero. 
The results of all three classification methods were similar. 
Prediction performance of Cox models indicated that logistic regression with ridge penalty and top-1000-variance genes outperformed the other two classification methods, while random forest tended to perform best in combination with all genes and with prognostic genes. 
However, the prediction performance of all Cox models was mainly moderate and not much better than random prediction. 
The combined model and the weighted model with fixed weights of increasing size showed the highest predictive accuracy, while the
estimated weights approach and standard subgroup model performed similarly badly. 
Genes identified most frequently by the former models were often present in all subgroups and some of them were reported in the literature to be associated with prognosis in various cancers. 
However, the corresponding estimated regression coefficients were often relatively small suggesting weak effects on survival outcome.
Few candidate genes with reported cancer relation and relatively strong subgroup-specific effects were selected most frequently by either the subgroup model or the proposed weighted model. 

A major reason for the overall moderate prediction accuracy in the application example may be that the present lung cancer studies are too heterogeneous. 
On the one hand, they comprise different histological subtypes that are known to be associated with a different prognosis.
One could think of using only patients belonging to the same histological subtype such as adenocarcinoma.
However, this would make the sizes of the patient subgroups even smaller.
On the other hand, tissue processing and RNA extraction for generating gene expression data as well as patient inclusion criteria vary between studies.
In GSE29013 genome-wide expression profiling was based on formalin-fixed paraffin-embedded (FFPE) tissues rather than fresh frozen tissues like in GSE37745 and GSE50081, which might influence expression levels.
GSE31210 and GSE50081 include only patients with stage I and II, and GSE31210 is additionally restricted to lung adenocarcinomas.

In \cite{madjar2018} we studied the influence of further parameters for weights estimation on prediction performance: the inclusion of interactions between genomic covariates and survival time in the classification model, as well as replacement of the survival time by the Nelson–Aalen estimator of the cumulative hazard rate (HR) in the set of covariates in the classification model. 
The latter was proposed by \cite{white2009} in the context of multiple imputation.
We also considered a simulation with unbalanced subgroup sizes and compared standard classification without sampling techniques with two oversampling techniques (random oversampling and synthetic minority oversampling technique). 
Oversampling increases the sample size of the small subgroup so that it is balanced with respect to the other subgroups. 
We found no considerable influence of the further parameters for weights estimation on prediction performance 
and also oversampling seemed to have no effect. 

\section*{Supplementary Materials}
Additional supporting information referenced in Sections~\ref{s:simres} and~\ref{s:applLC} are available with this paper.
The preprocessed lung cancer data analyzed in Section~\ref{s:applLC} and the R code implementing our method are available on GitHub: \\
{\small \url{https://github.com/KatrinMadjar/WeightedCoxRegression.git}}.


\bibliographystyle{unsrt}

\begin{thebibliography}{1}

\bibitem{bender2005}
Bender, R., Augustin, T., and Blettner, M. (2005).
\newblock Generating survival times to simulate {Cox} proportional hazards
  models.
\newblock {\em Statistics in Medicine} {\bf 24,} 1713--1723.

\bibitem{bergersen2011}
Bergersen, L.~C., Glad, I.~K., and Lyng, H. (2011).
\newblock Weighted lasso with data integration.
\newblock {\em Statistical Applications in Genetics and Molecular Biology} {\bf
  10,}.

\bibitem{bickel2008}
Bickel, S., Bogojeska, J., Lengauer, T., and Scheffer, T. (2008).
\newblock Multi-task {Learning} for {HIV} {Therapy} {Screening}.
\newblock In {\em Proceedings of the 25th {International} {Conference} on
  {Machine} {Learning}}, {ICML} '08, pages 56--63, New York, USA. ACM.

\bibitem{binder2012}
Binder, H., Müller, T., Schwender, H., Golka, K., Steffens, M., Hengstler,
  J.~G., Ickstadt, K., and Schumacher, M. (2012).
\newblock Cluster-localized sparse logistic regression for {SNP} data.
\newblock {\em Statistical Applications in Genetics and Molecular Biology} {\bf
  11,}.

\bibitem{binder2008}
Binder, H. and Schumacher, M. (2008).
\newblock Allowing for mandatory covariates in boosting estimation of sparse
  high-dimensional survival models.
\newblock {\em BMC Bioinformatics} {\bf 9,} 14.

\bibitem{bischl2016}
Bischl, B., Lang, M., Kotthoff, L., Schiffner, J., Richter, J., Studerus, E.,
  Casalicchio, G., and Jones, Z.~M. (2016).
\newblock {mlr}: Machine learning in r.
\newblock {\em Journal of Machine Learning Research} {\bf 17,} 1--5.

\bibitem{bogojeska2012}
Bogojeska, J. and Lengauer, T. (2012).
\newblock Hierarchical {Bayes} {Model} for {Predicting} {Effectiveness} of
  {HIV} {Combination} {Therapies}.
\newblock {\em Statistical Applications in Genetics and Molecular Biology} {\bf
  11,}.

\bibitem{boulesteix2017}
Boulesteix, A.-L., De~Bin, R., Jiang, X., and Fuchs, M. (2017).
\newblock {IPF}-{LASSO}: {Integrative} $l_1$-{Penalized} {Regression} with
  {Penalty} {Factors} for {Prediction} {Based} on {Multi}-{Omics} {Data}.
\newblock {\em Computational and Mathematical Methods in Medicine.} Vol. 2017,
  Article ID 7691937, 14 pages.

\bibitem{brown2015}
Brown, G.~R., Hem, V., Katz, K.~S., Ovetsky, M., Wallin, C., Ermolaeva, O.,
  Tolstoy, I., Tatusova, T., Pruitt, K.~D., Maglott, D.~R., and Murphy, T.~D.
  (2015).
\newblock Gene: a gene-centered information resource at {NCBI}.
\newblock {\em Nucleic Acids Research} {\bf 43,} D36--D42.

\bibitem{carlson2016}
Carlson, M. (2016).
\newblock {\em hgu133plus2.db: Affymetrix Human Genome U133 Plus 2.0 Array
  annotation data (chip hgu133plus2)}.
\newblock R package version 3.2.3.

\bibitem{cox_regression_1972}
Cox, D.~R. (1972).
\newblock Regression {Models} and {Life}-{Tables}.
\newblock {\em Journal of the Royal Statistical Society. Series B
  (Methodological)} {\bf 34,} 187--220.

\bibitem{edgar2002}
Edgar, R., Domrachev, M., and Lash, A.~E. (2002).
\newblock Gene {Expression} {Omnibus}: {NCBI} gene expression and hybridization
  array data repository.
\newblock {\em Nucleic Acids Research} {\bf 30,} 207--210.

\bibitem{friedman2010}
Friedman, J., Hastie, T., and Tibshirani, R. (2010).
\newblock Regularization {Paths} for {Generalized} {Linear} {Models} via
  {Coordinate} {Descent}.
\newblock {\em Journal of Statistical Software} {\bf 33,} 1--22.

\bibitem{gade2011}
Gade, S., Porzelius, C., Fälth, M., Brase, J.~C., Wuttig, D., Kuner, R.,
  Binder, H., Sültmann, H., and Beißbarth, T. (2011).
\newblock Graph based fusion of {miRNA} and {mRNA} expression data improves
  clinical outcome prediction in prostate cancer.
\newblock {\em BMC Bioinformatics} {\bf 12,} 488.

\bibitem{GeneCards2018}
GeneCards.
\newblock Genecards\textsuperscript{\textregistered}: The human gene database.
\newblock \url{https://www.genecards.org}.
\newblock Accessed: June 2018.

\bibitem{harrell1996}
Harrell, F.~E., Lee, K.~L., and Mark, D.~B. (1996).
\newblock Multivariable {Prognostic} {Models}: {Issues} in {Developing}
  {Models}, {Evaluating} {Assumptions} and {Adequacy}, and {Measuring} and
  {Reducing} {Errors}.
\newblock {\em Statistics in Medicine} {\bf 15,} 361--387.

\bibitem{harrell2018}
{Harrell Jr}, F.~E., with contributions~from Charles~Dupont, and many others.
  (2018).
\newblock {\em Hmisc: Harrell Miscellaneous}.
\newblock R package version 4.1-1.

\bibitem{hellwig2016}
Hellwig, B., Madjar, K., Edlund, K., Marchan, R., Cadenas, C., Heimes, A.-S.,
  Almstedt, K., Lebrecht, A., Sicking, I., Battista, M.~J., Micke, P., Schmidt,
  M., Hengstler, J.~G., and Rahnenführer, J. (2016).
\newblock Epsin {Family} {Member} 3 and {Ribosome}-{Related} {Genes} {Are}
  {Associated} with {Late} {Metastasis} in {Estrogen} {Receptor}-{Positive}
  {Breast} {Cancer} and {Long}-{Term} {Survival} in {Non}-{Small} {Cell} {Lung}
  {Cancer} {Using} a {Genome}-{Wide} {Identification} and {Validation}
  {Strategy}.
\newblock {\em PLOS ONE} {\bf 11,} e0167585.

\bibitem{hothorn2006a}
Hothorn, T. and Bühlmann, P. (2006).
\newblock Model-based boosting in high dimensions.
\newblock {\em Bioinformatics} {\bf 22,} 2828--2829.

\bibitem{hothorn2006b}
Hothorn, T., Bühlmann, P., Dudoit, S., Molinaro, A., and Van Der~Laan, M.~J.
  (2006).
\newblock Survival ensembles.
\newblock {\em Biostatistics} {\bf 7,} 355--373.

\bibitem{kratz2012}
Kratz, J.~R., He, J., Van Den~Eeden, S.~K., Zhu, Z.-H., Gao, W., Pham, P.~T.,
  Mulvihill, M.~S., Ziaei, F., Zhang, H., Su, B., Zhi, X., Quesenberry, C.~P.,
  Habel, L.~A., Deng, Q., Wang, Z., Zhou, J., Li, H., Huang, M.-C., Yeh, C.-C.,
  Segal, M.~R., Ray, M.~R., Jones, K.~D., Raz, D.~J., Xu, Z., Jahan, T.~M.,
  Berryman, D., He, B., Mann, M.~J., and Jablons, D.~M. (2012).
\newblock A practical molecular assay to predict survival in resected
  non-squamous, non-small-cell lung cancer: development and international
  validation studies.
\newblock {\em The Lancet} {\bf 379,} 823--832.

\bibitem{lang2017}
Lang, M., Bischl, B., and Surmann, D. (2017).
\newblock batchtools: Tools for r to work on batch systems.
\newblock {\em The Journal of Open Source Software} {\bf 2,}.

\bibitem{liu2014b}
Liu, J., Huang, J., and Ma, S. (2014).
\newblock Integrative {Analysis} of {Cancer} {Diagnosis} {Studies} with
  {Composite} {Penalization}.
\newblock {\em Scandinavian Journal of Statistics, Theory and Applications}
  {\bf 41,} 87--103.

\bibitem{liu2014}
Liu, J., Huang, J., Zhang, Y., Lan, Q., Rothman, N., Zheng, T., and Ma, S.
  (2014).
\newblock Integrative {Analysis} of {Prognosis} {Data} on {Multiple} {Cancer}
  {Subtypes}.
\newblock {\em Biometrics} {\bf 70,} 480--488.

\bibitem{madjar2018}
Madjar, K. (2018).
\newblock {\em Survival models with selection of genomic covariates in
  heterogeneous cancer studies}.
\newblock Dissertation. Faculty of Statistics, TU Dortmund University.

\bibitem{mccall2010}
McCall, M.~N., Bolstad, B.~M., and Irizarry, R.~A. (2010).
\newblock Frozen robust multiarray analysis ({fRMA}).
\newblock {\em Biostatistics} {\bf 11,} 242--253.

\bibitem{Richter2019}
Richter, J., Madjar, K., and Rahnenführer, J. (2019).
\newblock {Model-based optimization of subgroup weights for survival analysis}.
\newblock {\em Bioinformatics} {\bf 35,} i484--i491.

\bibitem{simon2002}
Simon, R. (2002).
\newblock Bayesian subset analysis: application to studying treatment-by-gender
  interactions.
\newblock {\em Statistics in Medicine} {\bf 21,} 2909--2916.

\bibitem{tang2017}
Tang, H., Wang, S., Xiao, G., Schiller, J., Papadimitrakopoulou, V., Minna, J.,
  Wistuba, I.~I., and Xie, Y. (2017).
\newblock Comprehensive evaluation of published gene expression prognostic
  signatures for biomarker-based lung cancer clinical studies.
\newblock {\em Annals of Oncology} {\bf 28,} 733--740.

\bibitem{tibshirani1996}
Tibshirani, R. (1996).
\newblock Regression {Shrinkage} and {Selection} via the {Lasso}.
\newblock {\em Journal of the Royal Statistical Society. Series B
  (Methodological)} {\bf 58,} 267--288.

\bibitem{tibshirani1997}
Tibshirani, R. (1997).
\newblock The {Lasso} {Method} for {Variable} {Selection} in the {Cox} {Model}.
\newblock {\em Statistics in Medicine} {\bf 16,} 385--395.

\bibitem{tutz2005}
Tutz, G. and Binder, H. (2005).
\newblock Localized classification.
\newblock {\em Statistics and Computing} {\bf 15,} 155--166.

\bibitem{tutz2006}
Tutz, G. and Binder, H. (2006).
\newblock Generalized {Additive} {Modeling} with {Implicit} {Variable}
  {Selection} by {Likelihood}-{Based} {Boosting}.
\newblock {\em Biometrics} {\bf 62,} 961--971.

\bibitem{verweij1994}
Verweij, P.~J. and Van~Houwelingen, H.~C. (1994).
\newblock Penalized likelihood in {Cox} regression.
\newblock {\em Statistics in Medicine} {\bf 13,} 2427--2436.

\bibitem{weyer2015}
Weyer, V. and Binder, H. (2015).
\newblock A weighting approach for judging the effect of patient strata on
  high-dimensional risk prediction signatures.
\newblock {\em BMC Bioinformatics} {\bf 16,} 294.

\bibitem{white2009}
White, I.~R. and Royston, P. (2009).
\newblock Imputing missing covariate values for the {Cox} model.
\newblock {\em Statistics in Medicine} {\bf 28,} 1982--1998.

\bibitem{wright2017}
Wright, M.~N. and Ziegler, A. (2017).
\newblock {ranger}: A fast implementation of random forests for high
  dimensional data in {C++} and {R}.
\newblock {\em Journal of Statistical Software} {\bf 77,} 1--17.

\bibitem{zerbino2018}
Zerbino, D.~R., Achuthan, P., Akanni, W., Amode, M.~R., Barrell, D., Bhai, J.,
  and {51 other authors} (2018).
\newblock Ensembl 2018.
\newblock {\em Nucleic Acids Research} {\bf 46,} D754--D761.

\bibitem{zou2005}
Zou, H. and Hastie, T. (2005).
\newblock Regularization and variable selection via the elastic net.
\newblock {\em Journal of the Royal Statistical Society: Series B (Statistical
  Methodology)} {\bf 67,} 301--320.

\end{thebibliography}


\appendix

\section*{\centering Supplementary Materials}

\subsection*{Preprocessing of NSCLC cohorts}

Four non-small cell lung cancer (NSCLC) cohorts with overall survival and censoring information, clinical pathologic information, and Affymetrix gene expression data of the tumor material, were downloaded from the Gene Expression Omnibus (GEO) data repository \cite{edgar2002} and manually curated as follows.
Raw gene expression data (CEL-files), measured on the Affymetrix HG-U133 Plus 2.0 array, were normalized using frozen robust multiarray analysis (fRMA) \cite{mccall2010}.
All cohorts were checked for duplicates by looking at correlations of the expression value vectors. 
Duplicates, small cell cancer samples and normal (non-tumorous) samples, as well as samples with missing survival endpoint were removed. 
More details on the data curation process can be found in \cite{hellwig2016}.
The resulting four NSCLC cohorts comprise $n=635$ patients with available overall survival endpoint, clinical variables (age at time of diagnosis, sex, pTNM stage, histology and smoking status), and gene expression data: GSE29013 ($n=55$, 18 events), GSE31210 ($n=226$, 35 events), GSE37745 ($n=194$, 143 events), and GSE50081 ($n=160$, 65 events). 
A summary of clinical pathologic variables of all cohorts is presented in Table~\ref{tab_clin_nsclc}.

The total number of measured genetic covariates (probe sets representing genes) in each cohort is 54675. 
For the majority of these probe sets the expression values are noisy and do not contain relevant information regarding survival outcome. 
This makes the identification of the prognostic genes more difficult and increases computation time.
Therefore, besides, we use two preselected reduced gene sets for analysis. 
One subset is defined by the 1000 features (probe sets) with the highest variance in gene expression values across all four cohorts, referred to as top-1000-variance genes.
This selection is based on the assumption that important prognostic genes imply systematic changes in their expression values and thus, a larger variance compared to irrelevant noise genes.
The second subset is a literature-based selection of 14 prognostic genes from \cite{kratz2012} and 20 prognostic gene signatures from a systematic literature review and meta-analysis-based evaluation by \cite{tang2017}. 
The review by \cite{tang2017} includes 42 published gene signatures derived from genome-wide mRNA gene expression studies, whereof 17 and 8 signatures, respectively, were found to be prognostic in the histological types adenocarcinoma and squamous cell carcinoma (a total of 20 different signatures as 5 signatures are prognostic in both histological types). 
We ignore how the genes were combined numerically in the original signatures (using statistical models) and combine only single genes with the 14 genes from \cite{kratz2012} to one prognostic gene list.
We translate gene symbols into corresponding probe set IDs of the Affymetrix HG-U133 Plus 2.0 array using the R/Bioconductor annotation package hgu133plus2.db \cite{carlson2016}. 
Not all genes have a match on this array.
Therefore, we use a reduced prognostic gene list for analysis comprising 3429 different probe sets that are related to 1323 different genes.

\newpage
\subsection*{Supplementary Figures}

\begin{figure}[!htb] 
\centering
\includegraphics[width=.9\textwidth]{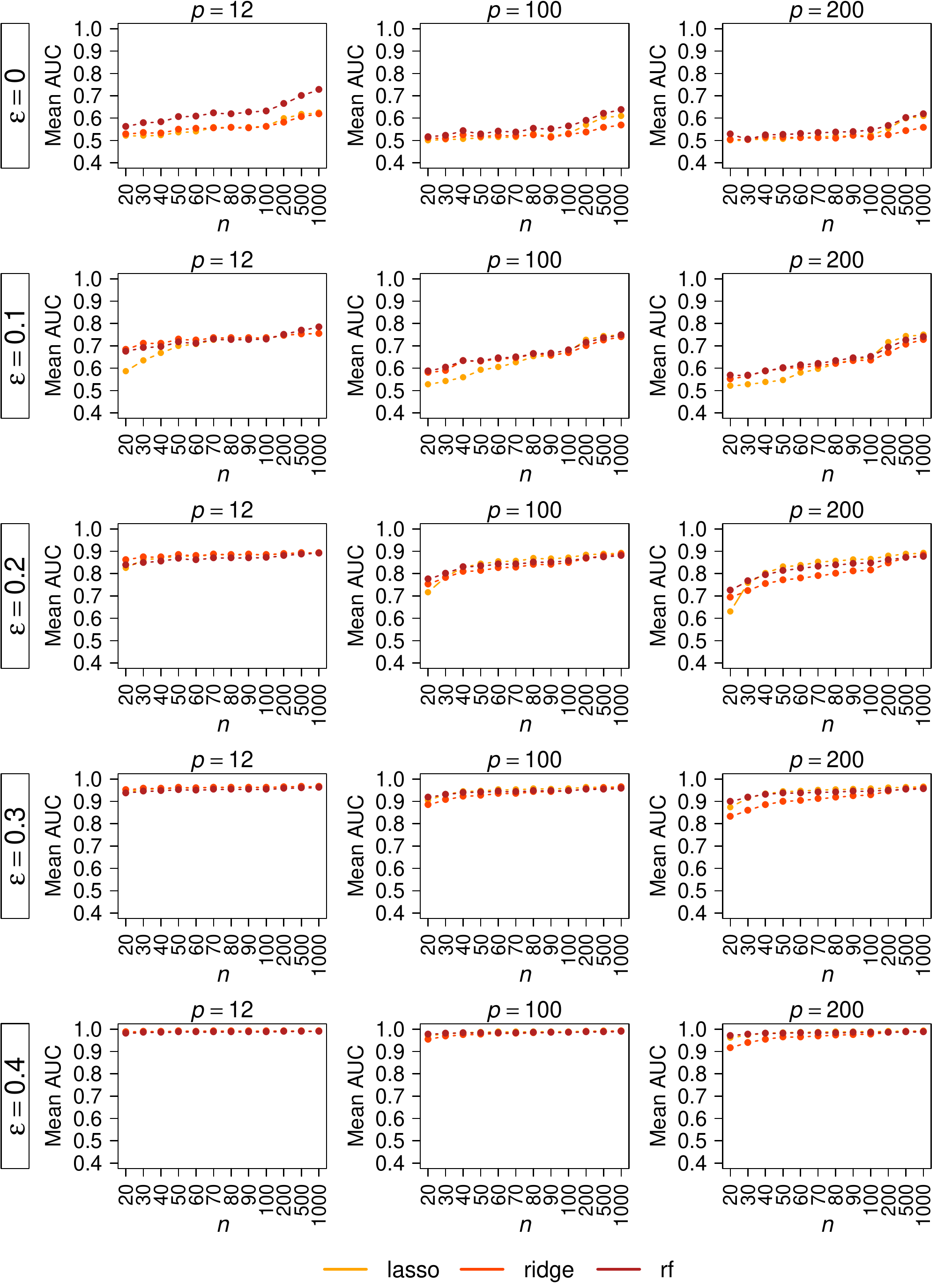}
\caption{Mean AUC (averaged across all test sets) for the different classification methods (colors) and simulation scenarios.} 
\label{fig_sim_auctest}
\end{figure}

\begin{figure}[!htb] 
\centering
\includegraphics[page=6, width=1\textwidth, trim={0 0 0 1.1cm},clip]{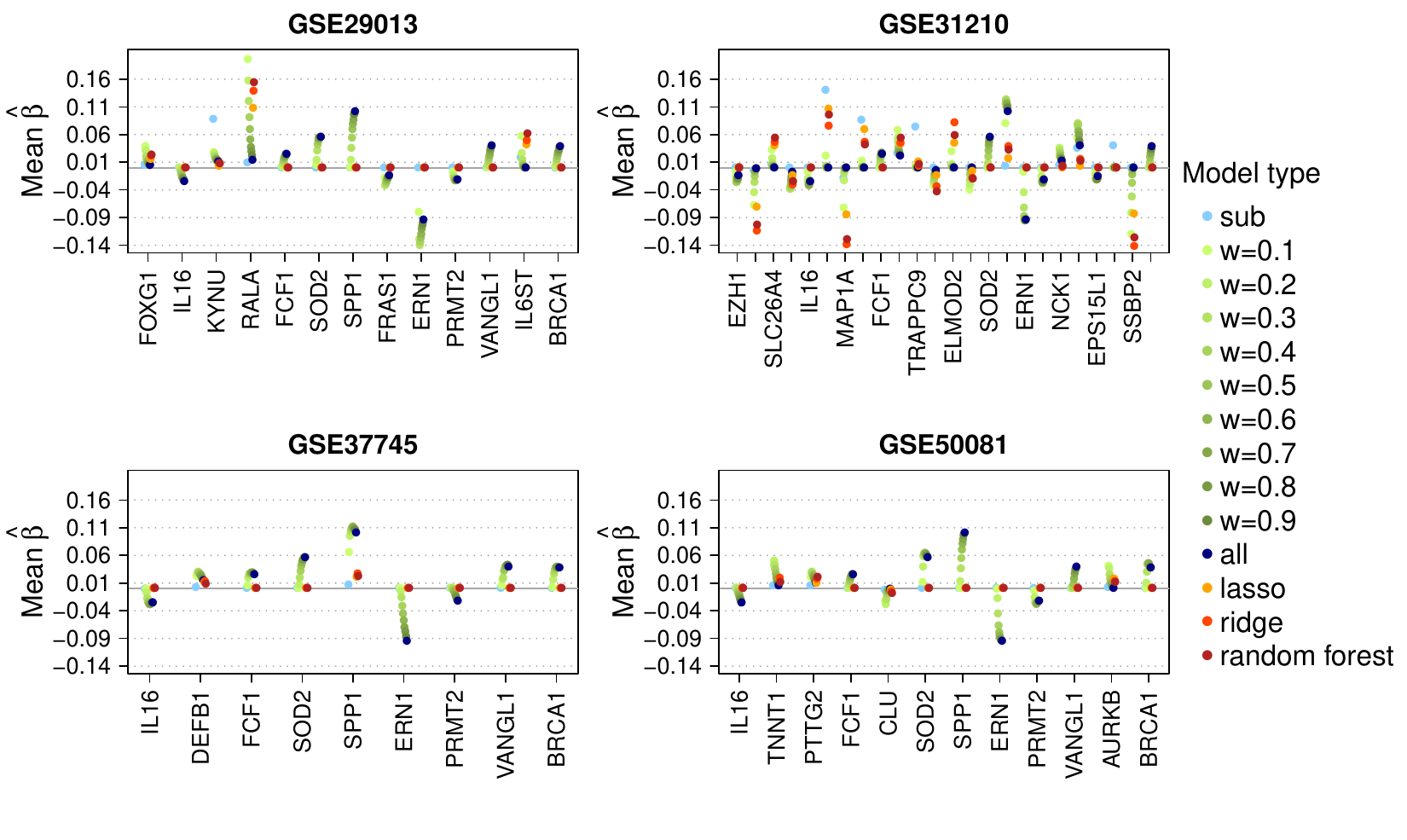}
\caption{Mean estimated regression coefficients (averaged across all training sets) of selected genes in all Cox models including the prognostic genes as covariates. For each subgroup genes with a mean inclusion frequency larger than 0.4 in any model type are selected.} \label{fig_LC_betas2}
\end{figure}

\begin{figure}[!htb] 
\centering
\includegraphics[page=5, width=1\textwidth, trim={0 0 0 1.1cm},clip]{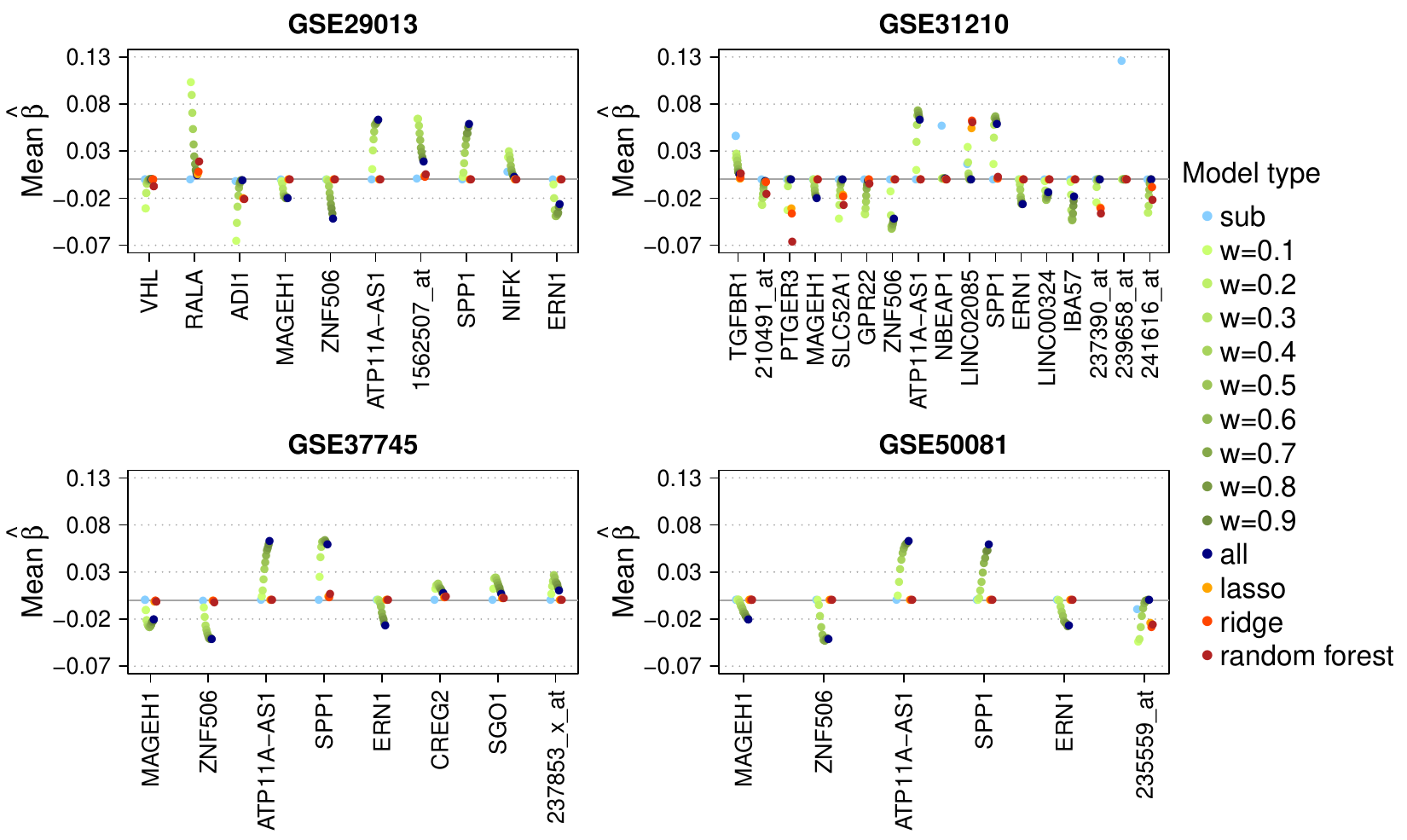}
\caption{Mean estimated regression coefficients (averaged across all training sets) of selected genes in all Cox models including all genes as covariates. For each subgroup genes with a mean inclusion frequency larger than 0.4 in any model type are selected.} 
\label{fig_LC_betas3}
\end{figure}

\begin{figure}[!htb] 
\centering
\includegraphics[page=1, width=.45\textwidth]{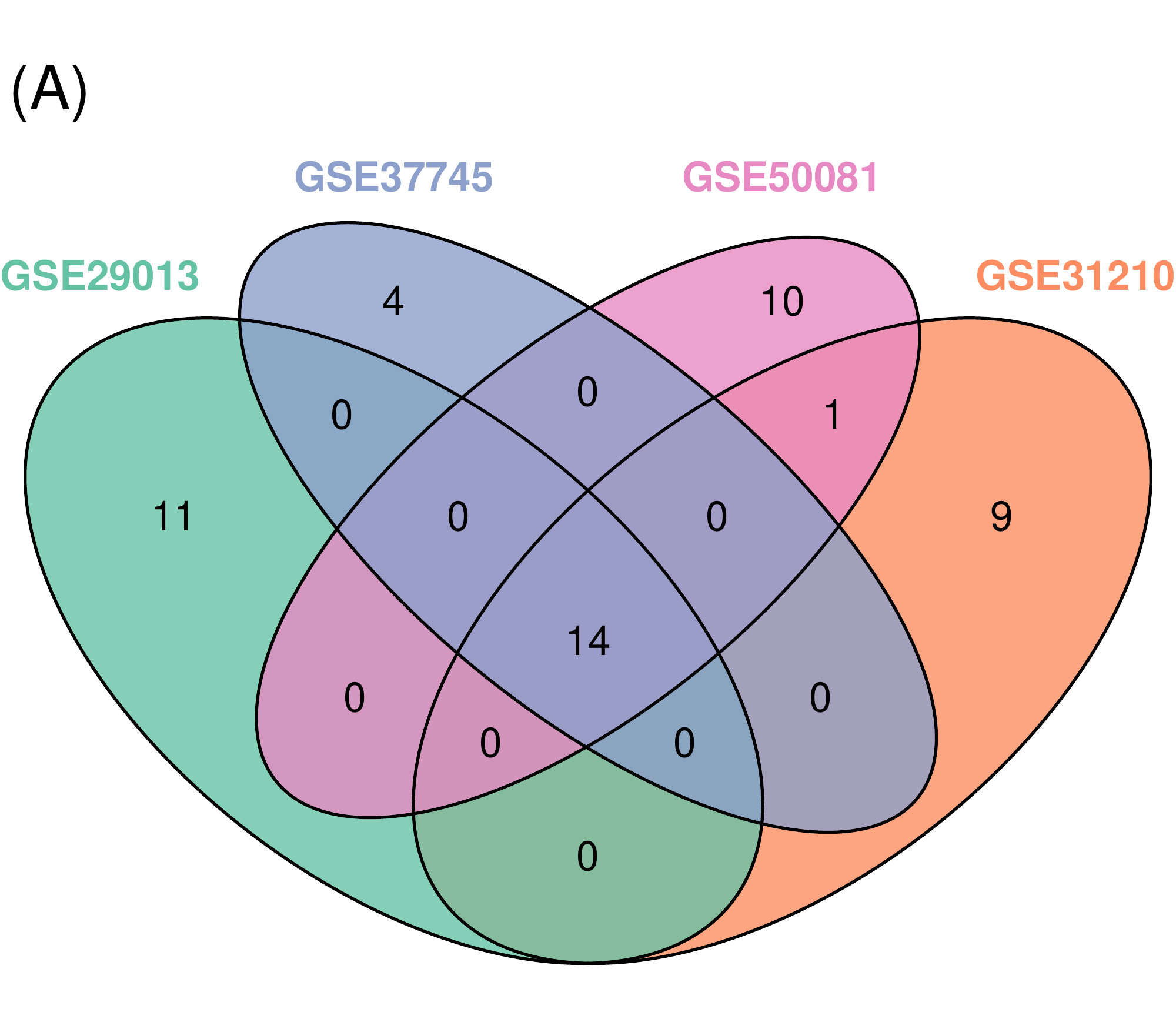} \hspace{.5cm}
\includegraphics[page=2, width=.45\textwidth]{VennDiagram_selGenes}
\includegraphics[page=3, width=.45\textwidth]{VennDiagram_selGenes}
\caption{Venn diagram of the number of selected genes in all subgroup Cox models including as covariates the (A) top-1000-variance genes, (B) prognostic genes, (C) all genes.
For each gene set and subgroup, genes with a mean inclusion frequency larger than 0.4 in any model type are selected.} 
\label{fig_LC_venn}
\end{figure}

\begin{figure}[!htb] 
\centering
\includegraphics[page=6, width=1\textwidth, trim={0 0 0 1cm},clip]{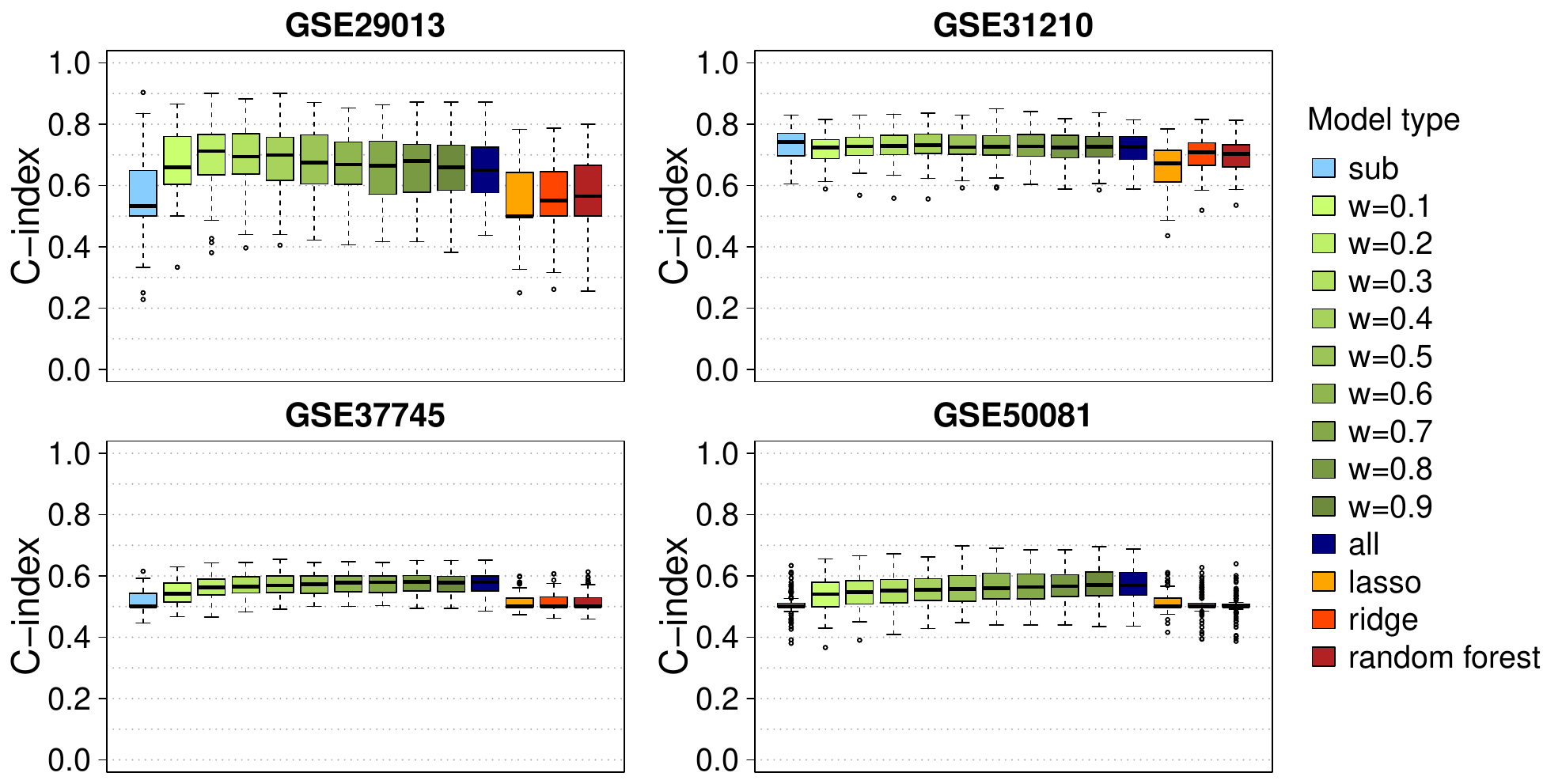}
\caption{Boxplots of C-index based on all test sets for the prediction of each subgroup under different types of Cox models including the prognostic genes.}
\label{fig_LC_CI2}
\end{figure}

\begin{figure}[!htb] 
\centering
\includegraphics[page=5, width=1\textwidth, trim={0 0 0 1cm},clip]{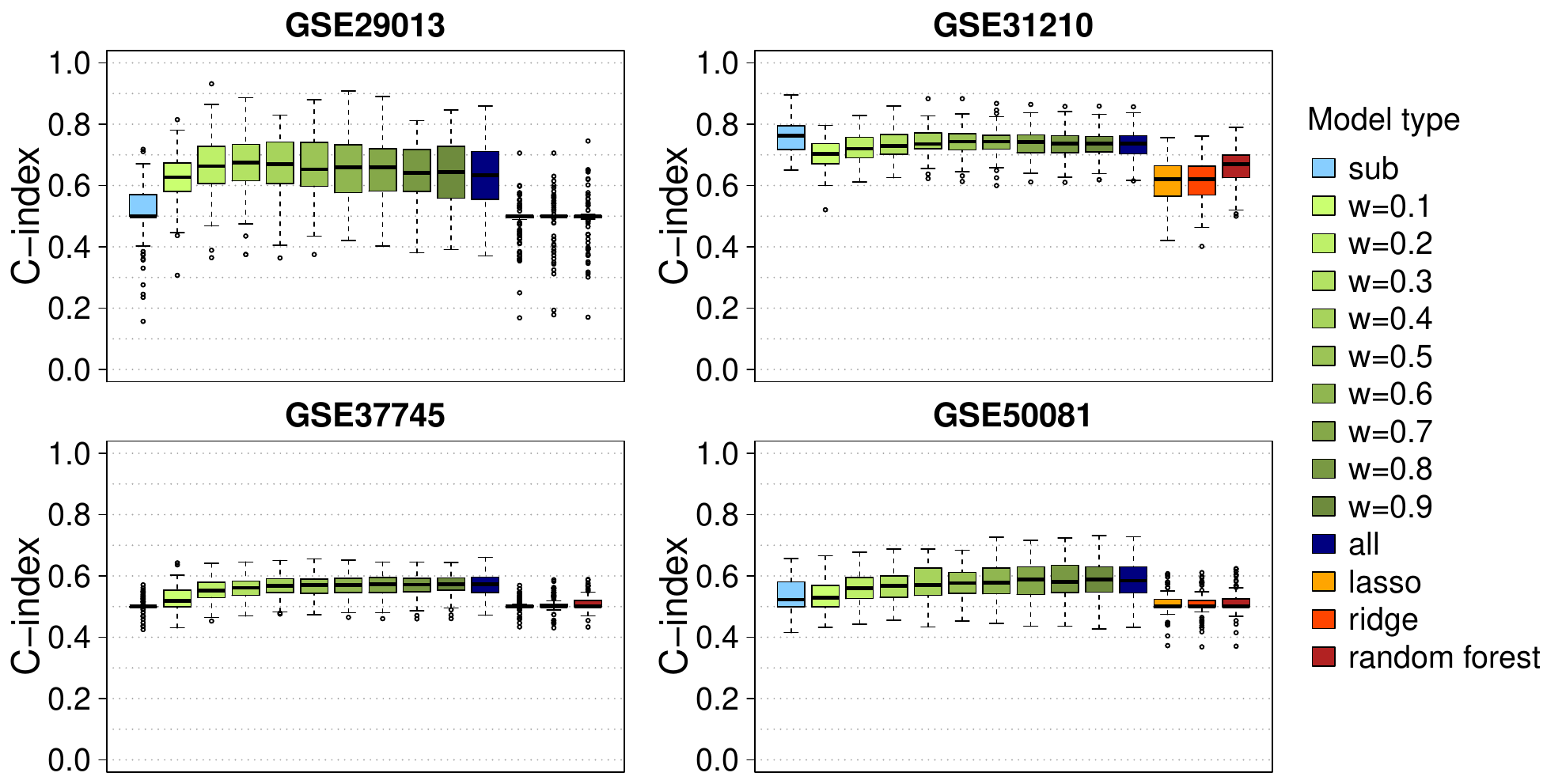}
\caption{Boxplots of C-index based on all test sets for the prediction of each subgroup under different types of Cox models including all genes.}
\label{fig_LC_CI3}
\end{figure} 
\clearpage
\subsection*{Supplementary Tables}

\begin{table}[!htb]
\caption{Summary of clinical pathologic variables of all NSCLC cohorts. Absolute frequencies of variable values.} 
\label{tab_clin_nsclc}
\begin{center}
\begin{tabular}{ll*{4}{r}}
\toprule
\rotatebox{66}{Variable} & \rotatebox{66}{Values} & \rotatebox{66}{GSE29013} & \rotatebox{66}{GSE31210} & \rotatebox{66}{GSE37745} & \rotatebox{66}{GSE50081} \\
	\midrule
	Sample size &  & 55 & 226 & 194 & 160 \\ 
	\midrule 
\multirow{3}{*}{Age (years)} & min. & 32 & 30 & 39 & 40 \\ 
   & mean & 64 & 60 & 64 & 68 \\ 
   & max. & 76 & 76 & 84 & 87 \\ 
   \midrule
\multirow{2}{*}{Sex} & male & 38 & 105 & 105 & 88 \\ 
   & female & 17 & 121 & 89 & 72 \\ 
   \midrule
\multirow{2}{*}{pTNM stage} & I & 24 & 168 & 128 & 112 \\ 
   & II-IV & 31 & 58 & 66 & 48 \\ 
   \midrule
\multirow{3}{*}{Histology} & SQC & 25 & 0 & 64 & 35 \\ 
   & ADC & 30 & 226 & 106 & 115 \\ 
   & other NSCLC & 0 & 0 & 24 & 10 \\ 
   \midrule
\multirow{2}{*}{Smoking status} & never-smoker & 2 & 115 & 15 & 24 \\ 
   & current/ former smoker & 53 & 111 & 179 & 136 \\ 
   \midrule
\multirow{2}{*}{Survival status} & censoring & 37 & 191 & 51 & 95 \\ 
   & event & 18 & 35 & 143 & 65 \\ 
	\bottomrule
  \end{tabular}
	\end{center}
\end{table}

\clearpage
\begin{landscape}
\small
\begin{longtable}[htb]{ccrrrrrrrrrrrrrrr}
\captionsetup{width=.9\linewidth}
$\epsilon$ & $p$ & $n$ & sub & $w=0.1$ & $w=0.2$ & $w=0.3$ & $w=0.4$ & $w=0.5$ & $w=0.6$ & $w=0.7$ & $w=0.8$ & $w=0.9$ & all & lasso & ridge & rf \\ 
  \hline
 \multirow{12}{*}{0} & \multirow{12}{*}{12} &  20 & 0.68 & 0.75 & 0.76 & 0.77 & 0.76 & 0.76 & 0.76 & 0.75 & 0.75 & 0.75 & 0.75 & 0.74 & 0.74 & 0.74 \\* 
   &  &  30 & 0.76 & 0.81 & 0.81 & 0.81 & 0.81 & 0.80 & 0.79 & 0.79 & 0.79 & 0.78 & 0.78 & 0.78 & 0.78 & 0.78 \\* 
   &  &  40 & 0.80 & 0.83 & 0.82 & 0.81 & 0.81 & 0.80 & 0.80 & 0.79 & 0.79 & 0.79 & 0.78 & 0.79 & 0.78 & 0.79 \\* 
   &  &  50 & 0.82 & 0.84 & 0.83 & 0.82 & 0.81 & 0.81 & 0.80 & 0.80 & 0.79 & 0.79 & 0.78 & 0.78 & 0.78 & 0.79 \\* 
   &  &  60 & 0.84 & 0.84 & 0.83 & 0.82 & 0.82 & 0.81 & 0.81 & 0.80 & 0.80 & 0.79 & 0.79 & 0.79 & 0.79 & 0.80 \\* 
   &  &  70 & 0.85 & 0.85 & 0.84 & 0.83 & 0.82 & 0.81 & 0.81 & 0.80 & 0.80 & 0.80 & 0.79 & 0.79 & 0.79 & 0.80 \\* 
   &  &  80 & 0.85 & 0.85 & 0.84 & 0.83 & 0.82 & 0.81 & 0.81 & 0.80 & 0.80 & 0.80 & 0.79 & 0.79 & 0.79 & 0.80 \\* 
   &  &  90 & 0.85 & 0.85 & 0.84 & 0.83 & 0.82 & 0.81 & 0.81 & 0.80 & 0.80 & 0.79 & 0.79 & 0.79 & 0.79 & 0.80 \\* 
   &  & 100 & 0.86 & 0.85 & 0.84 & 0.83 & 0.83 & 0.82 & 0.81 & 0.81 & 0.80 & 0.80 & 0.80 & 0.80 & 0.80 & 0.81 \\* 
   &  & 200 & 0.87 & 0.86 & 0.85 & 0.84 & 0.83 & 0.82 & 0.82 & 0.81 & 0.81 & 0.80 & 0.80 & 0.80 & 0.80 & 0.82 \\* 
   &  & 500 & 0.87 & 0.86 & 0.85 & 0.84 & 0.83 & 0.82 & 0.82 & 0.81 & 0.81 & 0.80 & 0.80 & 0.81 & 0.80 & 0.83 \\* 
   &  & 1000 & 0.87 & 0.86 & 0.85 & 0.84 & 0.83 & 0.82 & 0.82 & 0.81 & 0.81 & 0.80 & 0.80 & 0.80 & 0.81 & 0.83 \\ 
   \hline
\multirow{12}{*}{0} & \multirow{12}{*}{100} &  20 & 0.56 & 0.62 & 0.64 & 0.66 & 0.67 & 0.67 & 0.68 & 0.68 & 0.67 & 0.67 & 0.67 & 0.67 & 0.67 & 0.67 \\* 
   &  &  30 & 0.59 & 0.67 & 0.71 & 0.72 & 0.73 & 0.73 & 0.73 & 0.73 & 0.72 & 0.72 & 0.72 & 0.72 & 0.72 & 0.72 \\* 
   &  &  40 & 0.64 & 0.73 & 0.76 & 0.76 & 0.77 & 0.76 & 0.76 & 0.76 & 0.76 & 0.75 & 0.75 & 0.75 & 0.75 & 0.75 \\* 
   &  &  50 & 0.69 & 0.77 & 0.78 & 0.78 & 0.78 & 0.77 & 0.77 & 0.77 & 0.76 & 0.76 & 0.76 & 0.76 & 0.76 & 0.76 \\* 
   &  &  60 & 0.73 & 0.79 & 0.80 & 0.80 & 0.80 & 0.79 & 0.79 & 0.78 & 0.78 & 0.78 & 0.77 & 0.77 & 0.77 & 0.77 \\* 
   &  &  70 & 0.77 & 0.81 & 0.81 & 0.81 & 0.80 & 0.80 & 0.79 & 0.79 & 0.78 & 0.78 & 0.78 & 0.78 & 0.78 & 0.78 \\* 
   &  &  80 & 0.79 & 0.81 & 0.81 & 0.81 & 0.80 & 0.80 & 0.79 & 0.79 & 0.78 & 0.78 & 0.78 & 0.78 & 0.78 & 0.78 \\* 
   &  &  90 & 0.81 & 0.82 & 0.82 & 0.81 & 0.81 & 0.80 & 0.80 & 0.79 & 0.79 & 0.78 & 0.78 & 0.78 & 0.78 & 0.78 \\* 
   &  & 100 & 0.82 & 0.83 & 0.82 & 0.82 & 0.81 & 0.80 & 0.80 & 0.79 & 0.79 & 0.79 & 0.78 & 0.78 & 0.78 & 0.79 \\* 
   &  & 200 & 0.85 & 0.85 & 0.84 & 0.83 & 0.82 & 0.81 & 0.81 & 0.80 & 0.80 & 0.80 & 0.79 & 0.79 & 0.79 & 0.80 \\* 
   &  & 500 & 0.87 & 0.86 & 0.85 & 0.84 & 0.83 & 0.82 & 0.81 & 0.81 & 0.80 & 0.80 & 0.80 & 0.80 & 0.80 & 0.80 \\* 
   &  & 1000 & 0.87 & 0.86 & 0.85 & 0.84 & 0.83 & 0.82 & 0.82 & 0.81 & 0.81 & 0.80 & 0.80 & 0.80 & 0.80 & 0.81 \\ 
   \hline
\multirow{12}{*}{0} & \multirow{12}{*}{200} &  20 & 0.53 & 0.57 & 0.60 & 0.62 & 0.62 & 0.63 & 0.64 & 0.64 & 0.64 & 0.64 & 0.64 & 0.64 & 0.64 & 0.64 \\* 
   &  &  30 & 0.57 & 0.64 & 0.67 & 0.69 & 0.69 & 0.70 & 0.70 & 0.70 & 0.70 & 0.69 & 0.70 & 0.69 & 0.69 & 0.69 \\* 
   &  &  40 & 0.59 & 0.70 & 0.73 & 0.74 & 0.75 & 0.75 & 0.74 & 0.74 & 0.74 & 0.74 & 0.74 & 0.73 & 0.74 & 0.73 \\* 
   &  &  50 & 0.63 & 0.73 & 0.76 & 0.76 & 0.76 & 0.76 & 0.76 & 0.76 & 0.75 & 0.75 & 0.75 & 0.75 & 0.75 & 0.75 \\* 
   &  &  60 & 0.68 & 0.76 & 0.77 & 0.78 & 0.77 & 0.77 & 0.77 & 0.76 & 0.76 & 0.76 & 0.76 & 0.76 & 0.76 & 0.76 \\* 
   &  &  70 & 0.73 & 0.79 & 0.80 & 0.79 & 0.79 & 0.79 & 0.78 & 0.78 & 0.77 & 0.77 & 0.77 & 0.77 & 0.77 & 0.77 \\* 
   &  &  80 & 0.76 & 0.80 & 0.80 & 0.80 & 0.80 & 0.79 & 0.79 & 0.78 & 0.78 & 0.78 & 0.77 & 0.77 & 0.77 & 0.77 \\* 
   &  &  90 & 0.78 & 0.81 & 0.81 & 0.81 & 0.80 & 0.80 & 0.79 & 0.79 & 0.78 & 0.78 & 0.78 & 0.78 & 0.78 & 0.78 \\* 
   &  & 100 & 0.80 & 0.82 & 0.82 & 0.81 & 0.81 & 0.80 & 0.80 & 0.79 & 0.79 & 0.79 & 0.78 & 0.78 & 0.78 & 0.78 \\* 
   &  & 200 & 0.85 & 0.85 & 0.84 & 0.83 & 0.82 & 0.81 & 0.81 & 0.80 & 0.80 & 0.80 & 0.79 & 0.80 & 0.79 & 0.80 \\* 
   &  & 500 & 0.87 & 0.86 & 0.85 & 0.84 & 0.83 & 0.82 & 0.81 & 0.81 & 0.80 & 0.80 & 0.80 & 0.80 & 0.80 & 0.80 \\* 
   &  & 1000 & 0.87 & 0.86 & 0.85 & 0.84 & 0.83 & 0.82 & 0.82 & 0.81 & 0.81 & 0.80 & 0.80 & 0.80 & 0.80 & 0.80 \\ 
   \hline
\multirow{12}{*}{0.1} & \multirow{12}{*}{12} &  20 & 0.69 & 0.76 & 0.77 & 0.77 & 0.76 & 0.76 & 0.76 & 0.76 & 0.75 & 0.75 & 0.74 & 0.75 & 0.75 & 0.75 \\* 
   &  &  30 & 0.75 & 0.81 & 0.82 & 0.81 & 0.81 & 0.80 & 0.80 & 0.79 & 0.79 & 0.78 & 0.78 & 0.79 & 0.79 & 0.80 \\* 
   &  &  40 & 0.80 & 0.83 & 0.83 & 0.82 & 0.81 & 0.81 & 0.80 & 0.80 & 0.79 & 0.79 & 0.78 & 0.79 & 0.80 & 0.80 \\* 
   &  &  50 & 0.82 & 0.84 & 0.83 & 0.82 & 0.81 & 0.81 & 0.80 & 0.80 & 0.79 & 0.79 & 0.78 & 0.80 & 0.80 & 0.81 \\* 
   &  &  60 & 0.84 & 0.84 & 0.84 & 0.83 & 0.82 & 0.81 & 0.81 & 0.80 & 0.80 & 0.79 & 0.79 & 0.81 & 0.81 & 0.81 \\* 
   &  &  70 & 0.85 & 0.85 & 0.84 & 0.83 & 0.82 & 0.82 & 0.81 & 0.80 & 0.80 & 0.79 & 0.79 & 0.81 & 0.81 & 0.82 \\* 
   &  &  80 & 0.85 & 0.85 & 0.84 & 0.84 & 0.83 & 0.82 & 0.81 & 0.81 & 0.80 & 0.80 & 0.79 & 0.82 & 0.82 & 0.82 \\* 
   &  &  90 & 0.85 & 0.85 & 0.84 & 0.83 & 0.83 & 0.82 & 0.81 & 0.80 & 0.80 & 0.80 & 0.79 & 0.82 & 0.82 & 0.82 \\* 
   &  & 100 & 0.86 & 0.86 & 0.85 & 0.84 & 0.83 & 0.82 & 0.82 & 0.81 & 0.81 & 0.80 & 0.80 & 0.82 & 0.82 & 0.82 \\* 
   &  & 200 & 0.87 & 0.87 & 0.86 & 0.85 & 0.84 & 0.83 & 0.82 & 0.81 & 0.81 & 0.80 & 0.80 & 0.83 & 0.83 & 0.83 \\* 
   &  & 500 & 0.87 & 0.87 & 0.86 & 0.85 & 0.84 & 0.83 & 0.82 & 0.82 & 0.81 & 0.81 & 0.80 & 0.84 & 0.84 & 0.84 \\* 
   &  & 1000 & 0.87 & 0.87 & 0.86 & 0.85 & 0.84 & 0.83 & 0.82 & 0.82 & 0.81 & 0.80 & 0.80 & 0.84 & 0.84 & 0.85 \\ 
   \hline
\multirow{12}{*}{0.1} & \multirow{12}{*}{100} &  20 & 0.55 & 0.62 & 0.65 & 0.66 & 0.66 & 0.67 & 0.67 & 0.67 & 0.68 & 0.67 & 0.67 & 0.67 & 0.67 & 0.67 \\* 
   &  &  30 & 0.59 & 0.66 & 0.71 & 0.72 & 0.73 & 0.73 & 0.73 & 0.73 & 0.72 & 0.72 & 0.72 & 0.72 & 0.72 & 0.72 \\* 
   &  &  40 & 0.64 & 0.73 & 0.76 & 0.76 & 0.77 & 0.76 & 0.76 & 0.76 & 0.75 & 0.75 & 0.75 & 0.75 & 0.75 & 0.75 \\* 
   &  &  50 & 0.69 & 0.77 & 0.78 & 0.78 & 0.78 & 0.77 & 0.77 & 0.77 & 0.76 & 0.76 & 0.76 & 0.76 & 0.76 & 0.76 \\* 
   &  &  60 & 0.73 & 0.79 & 0.80 & 0.80 & 0.79 & 0.79 & 0.79 & 0.78 & 0.78 & 0.77 & 0.77 & 0.78 & 0.78 & 0.78 \\* 
   &  &  70 & 0.77 & 0.81 & 0.81 & 0.81 & 0.80 & 0.80 & 0.79 & 0.79 & 0.78 & 0.78 & 0.78 & 0.78 & 0.78 & 0.78 \\* 
   &  &  80 & 0.79 & 0.81 & 0.81 & 0.81 & 0.80 & 0.80 & 0.79 & 0.79 & 0.78 & 0.78 & 0.78 & 0.78 & 0.78 & 0.78 \\* 
   &  &  90 & 0.81 & 0.82 & 0.82 & 0.81 & 0.81 & 0.80 & 0.80 & 0.79 & 0.79 & 0.78 & 0.78 & 0.79 & 0.79 & 0.79 \\* 
   &  & 100 & 0.82 & 0.83 & 0.83 & 0.82 & 0.81 & 0.81 & 0.80 & 0.79 & 0.79 & 0.79 & 0.78 & 0.79 & 0.79 & 0.79 \\* 
   &  & 200 & 0.85 & 0.85 & 0.84 & 0.83 & 0.82 & 0.82 & 0.81 & 0.81 & 0.80 & 0.80 & 0.79 & 0.81 & 0.81 & 0.81 \\* 
   &  & 500 & 0.87 & 0.86 & 0.85 & 0.84 & 0.83 & 0.82 & 0.82 & 0.81 & 0.81 & 0.80 & 0.80 & 0.82 & 0.82 & 0.81 \\* 
   &  & 1000 & 0.87 & 0.87 & 0.86 & 0.85 & 0.84 & 0.83 & 0.82 & 0.81 & 0.81 & 0.80 & 0.80 & 0.83 & 0.83 & 0.82 \\ 
   \hline
\multirow{12}{*}{0.1} & \multirow{12}{*}{200} &  20 & 0.54 & 0.57 & 0.60 & 0.62 & 0.62 & 0.63 & 0.64 & 0.64 & 0.64 & 0.64 & 0.64 & 0.64 & 0.64 & 0.63 \\* 
   &  &  30 & 0.56 & 0.63 & 0.67 & 0.69 & 0.69 & 0.70 & 0.70 & 0.70 & 0.70 & 0.69 & 0.69 & 0.69 & 0.69 & 0.69 \\* 
   &  &  40 & 0.60 & 0.69 & 0.73 & 0.74 & 0.74 & 0.74 & 0.74 & 0.74 & 0.74 & 0.74 & 0.73 & 0.73 & 0.73 & 0.73 \\* 
   &  &  50 & 0.63 & 0.73 & 0.76 & 0.76 & 0.76 & 0.76 & 0.76 & 0.76 & 0.75 & 0.75 & 0.75 & 0.75 & 0.75 & 0.75 \\* 
   &  &  60 & 0.68 & 0.76 & 0.77 & 0.78 & 0.77 & 0.77 & 0.77 & 0.76 & 0.76 & 0.76 & 0.75 & 0.76 & 0.76 & 0.76 \\* 
   &  &  70 & 0.73 & 0.79 & 0.80 & 0.79 & 0.79 & 0.79 & 0.78 & 0.78 & 0.77 & 0.77 & 0.77 & 0.77 & 0.77 & 0.77 \\* 
   &  &  80 & 0.76 & 0.80 & 0.80 & 0.80 & 0.80 & 0.79 & 0.79 & 0.78 & 0.78 & 0.77 & 0.77 & 0.77 & 0.77 & 0.77 \\* 
   &  &  90 & 0.78 & 0.81 & 0.81 & 0.81 & 0.80 & 0.80 & 0.79 & 0.79 & 0.78 & 0.78 & 0.77 & 0.78 & 0.78 & 0.78 \\* 
   &  & 100 & 0.80 & 0.82 & 0.82 & 0.81 & 0.81 & 0.80 & 0.80 & 0.79 & 0.79 & 0.78 & 0.78 & 0.79 & 0.79 & 0.79 \\* 
   &  & 200 & 0.85 & 0.85 & 0.84 & 0.83 & 0.82 & 0.82 & 0.81 & 0.81 & 0.80 & 0.80 & 0.79 & 0.81 & 0.80 & 0.80 \\* 
   &  & 500 & 0.87 & 0.86 & 0.85 & 0.84 & 0.83 & 0.82 & 0.82 & 0.81 & 0.81 & 0.80 & 0.80 & 0.82 & 0.81 & 0.81 \\* 
   &  & 1000 & 0.87 & 0.87 & 0.86 & 0.85 & 0.83 & 0.83 & 0.82 & 0.81 & 0.81 & 0.80 & 0.80 & 0.83 & 0.82 & 0.81 \\ 
   \hline
\multirow{12}{*}{0.2} & \multirow{12}{*}{12} &  20 & 0.69 & 0.75 & 0.76 & 0.76 & 0.76 & 0.76 & 0.76 & 0.75 & 0.75 & 0.75 & 0.74 & 0.77 & 0.77 & 0.77 \\* 
   &  &  30 & 0.75 & 0.81 & 0.81 & 0.81 & 0.80 & 0.80 & 0.79 & 0.79 & 0.79 & 0.78 & 0.78 & 0.82 & 0.82 & 0.81 \\* 
   &  &  40 & 0.80 & 0.83 & 0.82 & 0.81 & 0.81 & 0.80 & 0.80 & 0.79 & 0.79 & 0.79 & 0.78 & 0.83 & 0.83 & 0.82 \\* 
   &  &  50 & 0.82 & 0.84 & 0.83 & 0.82 & 0.81 & 0.80 & 0.80 & 0.79 & 0.79 & 0.79 & 0.78 & 0.83 & 0.83 & 0.82 \\* 
   &  &  60 & 0.84 & 0.84 & 0.83 & 0.83 & 0.82 & 0.81 & 0.80 & 0.80 & 0.80 & 0.79 & 0.79 & 0.84 & 0.84 & 0.83 \\* 
   &  &  70 & 0.85 & 0.85 & 0.84 & 0.83 & 0.82 & 0.81 & 0.81 & 0.80 & 0.80 & 0.79 & 0.79 & 0.85 & 0.85 & 0.84 \\* 
   &  &  80 & 0.85 & 0.85 & 0.84 & 0.83 & 0.82 & 0.82 & 0.81 & 0.80 & 0.80 & 0.80 & 0.79 & 0.85 & 0.85 & 0.84 \\* 
   &  &  90 & 0.85 & 0.85 & 0.84 & 0.83 & 0.82 & 0.81 & 0.81 & 0.80 & 0.80 & 0.79 & 0.79 & 0.85 & 0.85 & 0.84 \\* 
   &  & 100 & 0.86 & 0.86 & 0.85 & 0.84 & 0.83 & 0.82 & 0.81 & 0.81 & 0.80 & 0.80 & 0.80 & 0.85 & 0.85 & 0.84 \\* 
   &  & 200 & 0.87 & 0.86 & 0.85 & 0.84 & 0.83 & 0.82 & 0.82 & 0.81 & 0.81 & 0.80 & 0.80 & 0.86 & 0.86 & 0.85 \\* 
   &  & 500 & 0.87 & 0.87 & 0.86 & 0.85 & 0.84 & 0.83 & 0.82 & 0.81 & 0.81 & 0.80 & 0.80 & 0.87 & 0.87 & 0.86 \\* 
   &  & 1000 & 0.87 & 0.87 & 0.86 & 0.85 & 0.84 & 0.83 & 0.82 & 0.81 & 0.81 & 0.80 & 0.80 & 0.87 & 0.87 & 0.86 \\ 
   \hline
\multirow{12}{*}{0.2} & \multirow{12}{*}{100} &  20 & 0.56 & 0.61 & 0.65 & 0.65 & 0.66 & 0.67 & 0.67 & 0.67 & 0.67 & 0.67 & 0.67 & 0.67 & 0.67 & 0.67 \\* 
   &  &  30 & 0.59 & 0.66 & 0.70 & 0.71 & 0.72 & 0.72 & 0.73 & 0.72 & 0.72 & 0.72 & 0.72 & 0.72 & 0.72 & 0.72 \\* 
   &  &  40 & 0.64 & 0.72 & 0.75 & 0.76 & 0.76 & 0.76 & 0.76 & 0.75 & 0.75 & 0.75 & 0.75 & 0.76 & 0.76 & 0.76 \\* 
   &  &  50 & 0.68 & 0.76 & 0.77 & 0.77 & 0.77 & 0.77 & 0.77 & 0.76 & 0.76 & 0.76 & 0.76 & 0.78 & 0.77 & 0.77 \\* 
   &  &  60 & 0.73 & 0.79 & 0.79 & 0.79 & 0.79 & 0.78 & 0.78 & 0.78 & 0.78 & 0.77 & 0.77 & 0.80 & 0.79 & 0.79 \\* 
   &  &  70 & 0.77 & 0.80 & 0.80 & 0.80 & 0.79 & 0.79 & 0.79 & 0.78 & 0.78 & 0.78 & 0.77 & 0.81 & 0.80 & 0.79 \\* 
   &  &  80 & 0.79 & 0.81 & 0.81 & 0.80 & 0.80 & 0.79 & 0.79 & 0.78 & 0.78 & 0.78 & 0.77 & 0.81 & 0.80 & 0.79 \\* 
   &  &  90 & 0.81 & 0.82 & 0.81 & 0.81 & 0.80 & 0.79 & 0.79 & 0.79 & 0.78 & 0.78 & 0.78 & 0.82 & 0.81 & 0.80 \\* 
   &  & 100 & 0.82 & 0.83 & 0.82 & 0.81 & 0.80 & 0.80 & 0.79 & 0.79 & 0.79 & 0.78 & 0.78 & 0.82 & 0.81 & 0.80 \\* 
   &  & 200 & 0.85 & 0.85 & 0.84 & 0.83 & 0.82 & 0.81 & 0.81 & 0.80 & 0.80 & 0.79 & 0.79 & 0.84 & 0.83 & 0.82 \\* 
   &  & 500 & 0.87 & 0.86 & 0.85 & 0.84 & 0.83 & 0.82 & 0.81 & 0.81 & 0.80 & 0.80 & 0.80 & 0.86 & 0.85 & 0.83 \\* 
   &  & 1000 & 0.87 & 0.87 & 0.86 & 0.84 & 0.83 & 0.82 & 0.82 & 0.81 & 0.81 & 0.80 & 0.80 & 0.86 & 0.86 & 0.84 \\ 
   \hline
\multirow{12}{*}{0.2} & \multirow{12}{*}{200} &  20 & 0.53 & 0.57 & 0.59 & 0.61 & 0.62 & 0.63 & 0.63 & 0.64 & 0.63 & 0.64 & 0.63 & 0.63 & 0.64 & 0.64 \\* 
   &  &  30 & 0.56 & 0.63 & 0.66 & 0.68 & 0.69 & 0.69 & 0.70 & 0.69 & 0.69 & 0.70 & 0.69 & 0.69 & 0.69 & 0.70 \\* 
   &  &  40 & 0.60 & 0.68 & 0.72 & 0.73 & 0.74 & 0.74 & 0.74 & 0.74 & 0.73 & 0.73 & 0.73 & 0.74 & 0.73 & 0.74 \\* 
   &  &  50 & 0.64 & 0.72 & 0.74 & 0.75 & 0.75 & 0.75 & 0.75 & 0.75 & 0.75 & 0.75 & 0.75 & 0.76 & 0.75 & 0.75 \\* 
   &  &  60 & 0.68 & 0.75 & 0.76 & 0.77 & 0.77 & 0.76 & 0.76 & 0.76 & 0.76 & 0.76 & 0.75 & 0.77 & 0.76 & 0.76 \\* 
   &  &  70 & 0.73 & 0.78 & 0.79 & 0.79 & 0.78 & 0.78 & 0.78 & 0.78 & 0.77 & 0.77 & 0.77 & 0.79 & 0.78 & 0.78 \\* 
   &  &  80 & 0.76 & 0.79 & 0.80 & 0.79 & 0.79 & 0.78 & 0.78 & 0.78 & 0.77 & 0.77 & 0.77 & 0.80 & 0.79 & 0.78 \\* 
   &  &  90 & 0.78 & 0.81 & 0.80 & 0.80 & 0.79 & 0.79 & 0.78 & 0.78 & 0.78 & 0.78 & 0.77 & 0.81 & 0.79 & 0.79 \\* 
   &  & 100 & 0.80 & 0.81 & 0.81 & 0.81 & 0.80 & 0.79 & 0.79 & 0.79 & 0.78 & 0.78 & 0.78 & 0.81 & 0.80 & 0.80 \\* 
   &  & 200 & 0.85 & 0.85 & 0.84 & 0.83 & 0.82 & 0.81 & 0.81 & 0.80 & 0.80 & 0.80 & 0.79 & 0.84 & 0.82 & 0.81 \\* 
   &  & 500 & 0.87 & 0.86 & 0.85 & 0.84 & 0.83 & 0.82 & 0.81 & 0.81 & 0.80 & 0.80 & 0.80 & 0.85 & 0.84 & 0.82 \\* 
   &  & 1000 & 0.87 & 0.87 & 0.85 & 0.84 & 0.83 & 0.82 & 0.81 & 0.81 & 0.80 & 0.80 & 0.80 & 0.86 & 0.86 & 0.83 \\ 
   \hline
\multirow{12}{*}{0.3} & \multirow{12}{*}{12} &  20 & 0.68 & 0.74 & 0.75 & 0.76 & 0.75 & 0.76 & 0.75 & 0.75 & 0.75 & 0.75 & 0.75 & 0.78 & 0.79 & 0.78 \\* 
   &  &  30 & 0.75 & 0.80 & 0.80 & 0.80 & 0.80 & 0.79 & 0.79 & 0.79 & 0.79 & 0.78 & 0.78 & 0.83 & 0.83 & 0.82 \\* 
   &  &  40 & 0.80 & 0.82 & 0.82 & 0.81 & 0.80 & 0.80 & 0.80 & 0.79 & 0.79 & 0.79 & 0.79 & 0.84 & 0.84 & 0.83 \\* 
   &  &  50 & 0.82 & 0.83 & 0.82 & 0.81 & 0.80 & 0.80 & 0.79 & 0.79 & 0.79 & 0.78 & 0.78 & 0.85 & 0.85 & 0.84 \\* 
   &  &  60 & 0.84 & 0.84 & 0.83 & 0.82 & 0.81 & 0.81 & 0.80 & 0.80 & 0.79 & 0.79 & 0.79 & 0.85 & 0.85 & 0.84 \\* 
   &  &  70 & 0.85 & 0.84 & 0.83 & 0.82 & 0.81 & 0.81 & 0.80 & 0.80 & 0.80 & 0.79 & 0.79 & 0.86 & 0.86 & 0.85 \\* 
   &  &  80 & 0.85 & 0.85 & 0.84 & 0.83 & 0.82 & 0.81 & 0.81 & 0.80 & 0.80 & 0.79 & 0.79 & 0.86 & 0.86 & 0.85 \\* 
   &  &  90 & 0.85 & 0.85 & 0.84 & 0.82 & 0.82 & 0.81 & 0.80 & 0.80 & 0.80 & 0.79 & 0.79 & 0.86 & 0.86 & 0.85 \\* 
   &  & 100 & 0.86 & 0.85 & 0.84 & 0.83 & 0.82 & 0.81 & 0.81 & 0.81 & 0.80 & 0.80 & 0.80 & 0.86 & 0.86 & 0.86 \\* 
   &  & 200 & 0.87 & 0.86 & 0.85 & 0.84 & 0.83 & 0.82 & 0.81 & 0.81 & 0.81 & 0.80 & 0.80 & 0.87 & 0.87 & 0.86 \\* 
   &  & 500 & 0.87 & 0.86 & 0.85 & 0.84 & 0.83 & 0.82 & 0.82 & 0.81 & 0.81 & 0.80 & 0.80 & 0.87 & 0.87 & 0.87 \\* 
   &  & 1000 & 0.87 & 0.86 & 0.85 & 0.84 & 0.83 & 0.82 & 0.82 & 0.81 & 0.81 & 0.80 & 0.80 & 0.87 & 0.87 & 0.87 \\ 
   \hline
\multirow{12}{*}{0.3} & \multirow{12}{*}{100} &  20 & 0.56 & 0.61 & 0.63 & 0.65 & 0.66 & 0.67 & 0.67 & 0.67 & 0.67 & 0.67 & 0.66 & 0.65 & 0.67 & 0.67 \\* 
   &  &  30 & 0.59 & 0.65 & 0.69 & 0.71 & 0.72 & 0.72 & 0.72 & 0.72 & 0.72 & 0.72 & 0.72 & 0.72 & 0.72 & 0.73 \\* 
   &  &  40 & 0.65 & 0.71 & 0.74 & 0.75 & 0.75 & 0.75 & 0.75 & 0.75 & 0.75 & 0.75 & 0.75 & 0.77 & 0.76 & 0.76 \\* 
   &  &  50 & 0.69 & 0.74 & 0.76 & 0.76 & 0.76 & 0.76 & 0.76 & 0.76 & 0.76 & 0.76 & 0.75 & 0.79 & 0.78 & 0.77 \\* 
   &  &  60 & 0.73 & 0.77 & 0.78 & 0.78 & 0.78 & 0.78 & 0.78 & 0.77 & 0.77 & 0.77 & 0.77 & 0.81 & 0.80 & 0.79 \\* 
   &  &  70 & 0.77 & 0.79 & 0.79 & 0.79 & 0.79 & 0.78 & 0.78 & 0.78 & 0.78 & 0.77 & 0.77 & 0.82 & 0.81 & 0.80 \\* 
   &  &  80 & 0.79 & 0.80 & 0.80 & 0.79 & 0.79 & 0.78 & 0.78 & 0.78 & 0.78 & 0.78 & 0.77 & 0.83 & 0.81 & 0.80 \\* 
   &  &  90 & 0.81 & 0.81 & 0.80 & 0.80 & 0.79 & 0.79 & 0.79 & 0.78 & 0.78 & 0.78 & 0.78 & 0.83 & 0.82 & 0.81 \\* 
   &  & 100 & 0.82 & 0.82 & 0.81 & 0.80 & 0.80 & 0.79 & 0.79 & 0.79 & 0.79 & 0.78 & 0.78 & 0.84 & 0.82 & 0.81 \\* 
   &  & 200 & 0.85 & 0.84 & 0.83 & 0.82 & 0.81 & 0.81 & 0.80 & 0.80 & 0.80 & 0.79 & 0.79 & 0.86 & 0.85 & 0.83 \\* 
   &  & 500 & 0.87 & 0.86 & 0.84 & 0.83 & 0.82 & 0.81 & 0.81 & 0.80 & 0.80 & 0.80 & 0.80 & 0.87 & 0.87 & 0.85 \\* 
   &  & 1000 & 0.87 & 0.86 & 0.85 & 0.84 & 0.83 & 0.82 & 0.81 & 0.81 & 0.80 & 0.80 & 0.80 & 0.87 & 0.87 & 0.86 \\ 
   \hline
\multirow{12}{*}{0.3} & \multirow{12}{*}{200} &  20 & 0.54 & 0.57 & 0.59 & 0.60 & 0.62 & 0.63 & 0.63 & 0.63 & 0.63 & 0.64 & 0.63 & 0.63 & 0.63 & 0.63 \\* 
   &  &  30 & 0.56 & 0.62 & 0.65 & 0.67 & 0.68 & 0.69 & 0.69 & 0.69 & 0.69 & 0.69 & 0.69 & 0.69 & 0.70 & 0.70 \\* 
   &  &  40 & 0.60 & 0.67 & 0.71 & 0.72 & 0.73 & 0.73 & 0.73 & 0.73 & 0.73 & 0.73 & 0.73 & 0.74 & 0.73 & 0.74 \\* 
   &  &  50 & 0.63 & 0.70 & 0.73 & 0.74 & 0.75 & 0.75 & 0.75 & 0.75 & 0.75 & 0.74 & 0.74 & 0.77 & 0.75 & 0.76 \\* 
   &  &  60 & 0.68 & 0.73 & 0.75 & 0.76 & 0.76 & 0.76 & 0.76 & 0.76 & 0.75 & 0.75 & 0.75 & 0.78 & 0.76 & 0.77 \\* 
   &  &  70 & 0.73 & 0.77 & 0.78 & 0.78 & 0.78 & 0.78 & 0.77 & 0.77 & 0.77 & 0.77 & 0.77 & 0.81 & 0.79 & 0.78 \\* 
   &  &  80 & 0.76 & 0.78 & 0.78 & 0.78 & 0.78 & 0.78 & 0.77 & 0.77 & 0.77 & 0.77 & 0.77 & 0.81 & 0.79 & 0.79 \\* 
   &  &  90 & 0.78 & 0.79 & 0.79 & 0.79 & 0.78 & 0.78 & 0.78 & 0.78 & 0.78 & 0.77 & 0.77 & 0.82 & 0.80 & 0.79 \\* 
   &  & 100 & 0.80 & 0.80 & 0.80 & 0.80 & 0.79 & 0.79 & 0.79 & 0.78 & 0.78 & 0.78 & 0.78 & 0.83 & 0.81 & 0.80 \\* 
   &  & 200 & 0.85 & 0.84 & 0.83 & 0.82 & 0.81 & 0.81 & 0.80 & 0.80 & 0.80 & 0.79 & 0.79 & 0.85 & 0.84 & 0.82 \\* 
   &  & 500 & 0.87 & 0.86 & 0.84 & 0.83 & 0.82 & 0.81 & 0.81 & 0.80 & 0.80 & 0.80 & 0.80 & 0.87 & 0.86 & 0.84 \\* 
   &  & 1000 & 0.87 & 0.86 & 0.85 & 0.84 & 0.83 & 0.82 & 0.81 & 0.81 & 0.80 & 0.80 & 0.80 & 0.87 & 0.87 & 0.85 \\ 
   \hline
\multirow{12}{*}{0.4} & \multirow{12}{*}{12} &  20 & 0.68 & 0.74 & 0.74 & 0.75 & 0.75 & 0.75 & 0.75 & 0.75 & 0.75 & 0.75 & 0.75 & 0.78 & 0.79 & 0.78 \\* 
   &  &  30 & 0.75 & 0.79 & 0.80 & 0.80 & 0.79 & 0.79 & 0.79 & 0.79 & 0.79 & 0.79 & 0.78 & 0.83 & 0.83 & 0.83 \\* 
   &  &  40 & 0.80 & 0.81 & 0.81 & 0.80 & 0.80 & 0.80 & 0.79 & 0.79 & 0.79 & 0.79 & 0.79 & 0.85 & 0.85 & 0.84 \\* 
   &  &  50 & 0.82 & 0.82 & 0.81 & 0.81 & 0.80 & 0.80 & 0.79 & 0.79 & 0.79 & 0.78 & 0.78 & 0.85 & 0.85 & 0.84 \\* 
   &  &  60 & 0.84 & 0.83 & 0.82 & 0.81 & 0.81 & 0.80 & 0.80 & 0.80 & 0.79 & 0.79 & 0.79 & 0.86 & 0.86 & 0.85 \\* 
   &  &  70 & 0.85 & 0.84 & 0.82 & 0.82 & 0.81 & 0.81 & 0.80 & 0.80 & 0.79 & 0.79 & 0.79 & 0.86 & 0.86 & 0.85 \\* 
   &  &  80 & 0.85 & 0.84 & 0.83 & 0.82 & 0.81 & 0.81 & 0.80 & 0.80 & 0.80 & 0.79 & 0.79 & 0.86 & 0.86 & 0.86 \\* 
   &  &  90 & 0.85 & 0.84 & 0.83 & 0.82 & 0.81 & 0.81 & 0.80 & 0.80 & 0.80 & 0.79 & 0.79 & 0.86 & 0.86 & 0.86 \\* 
   &  & 100 & 0.86 & 0.85 & 0.83 & 0.82 & 0.82 & 0.81 & 0.81 & 0.80 & 0.80 & 0.80 & 0.80 & 0.87 & 0.87 & 0.86 \\* 
   &  & 200 & 0.87 & 0.86 & 0.84 & 0.83 & 0.82 & 0.82 & 0.81 & 0.81 & 0.80 & 0.80 & 0.80 & 0.87 & 0.87 & 0.87 \\* 
   &  & 500 & 0.87 & 0.86 & 0.84 & 0.83 & 0.83 & 0.82 & 0.81 & 0.81 & 0.81 & 0.80 & 0.80 & 0.87 & 0.87 & 0.87 \\* 
   &  & 1000 & 0.87 & 0.86 & 0.84 & 0.83 & 0.82 & 0.82 & 0.81 & 0.81 & 0.80 & 0.80 & 0.80 & 0.87 & 0.87 & 0.87 \\ 
   \hline
\multirow{12}{*}{0.4} & \multirow{12}{*}{100} &  20 & 0.55 & 0.60 & 0.62 & 0.64 & 0.65 & 0.66 & 0.66 & 0.66 & 0.67 & 0.67 & 0.66 & 0.65 & 0.67 & 0.66 \\* 
   &  &  30 & 0.59 & 0.64 & 0.68 & 0.70 & 0.71 & 0.71 & 0.71 & 0.71 & 0.71 & 0.71 & 0.71 & 0.71 & 0.72 & 0.73 \\* 
   &  &  40 & 0.64 & 0.70 & 0.73 & 0.74 & 0.74 & 0.75 & 0.75 & 0.75 & 0.75 & 0.74 & 0.74 & 0.76 & 0.76 & 0.76 \\* 
   &  &  50 & 0.69 & 0.73 & 0.75 & 0.75 & 0.75 & 0.76 & 0.76 & 0.75 & 0.75 & 0.75 & 0.75 & 0.79 & 0.77 & 0.77 \\* 
   &  &  60 & 0.73 & 0.76 & 0.77 & 0.77 & 0.77 & 0.77 & 0.77 & 0.77 & 0.77 & 0.77 & 0.77 & 0.81 & 0.80 & 0.79 \\* 
   &  &  70 & 0.77 & 0.77 & 0.78 & 0.78 & 0.78 & 0.78 & 0.78 & 0.78 & 0.77 & 0.77 & 0.77 & 0.82 & 0.81 & 0.80 \\* 
   &  &  80 & 0.79 & 0.78 & 0.78 & 0.78 & 0.78 & 0.78 & 0.78 & 0.78 & 0.78 & 0.77 & 0.77 & 0.83 & 0.81 & 0.80 \\* 
   &  &  90 & 0.81 & 0.80 & 0.79 & 0.79 & 0.79 & 0.79 & 0.78 & 0.78 & 0.78 & 0.78 & 0.78 & 0.84 & 0.82 & 0.81 \\* 
   &  & 100 & 0.82 & 0.80 & 0.80 & 0.80 & 0.79 & 0.79 & 0.79 & 0.79 & 0.78 & 0.78 & 0.78 & 0.84 & 0.83 & 0.82 \\* 
   &  & 200 & 0.85 & 0.84 & 0.82 & 0.81 & 0.81 & 0.80 & 0.80 & 0.80 & 0.79 & 0.79 & 0.79 & 0.86 & 0.85 & 0.84 \\* 
   &  & 500 & 0.87 & 0.85 & 0.84 & 0.82 & 0.82 & 0.81 & 0.81 & 0.80 & 0.80 & 0.80 & 0.80 & 0.87 & 0.87 & 0.85 \\* 
   &  & 1000 & 0.87 & 0.86 & 0.84 & 0.83 & 0.82 & 0.82 & 0.81 & 0.81 & 0.80 & 0.80 & 0.80 & 0.87 & 0.87 & 0.86 \\ 
   \hline
\multirow{12}{*}{0.4} & \multirow{12}{*}{200} &  20 & 0.54 & 0.56 & 0.58 & 0.60 & 0.61 & 0.62 & 0.62 & 0.63 & 0.63 & 0.63 & 0.63 & 0.62 & 0.63 & 0.63 \\* 
   &  &  30 & 0.56 & 0.62 & 0.65 & 0.67 & 0.68 & 0.69 & 0.69 & 0.69 & 0.69 & 0.69 & 0.69 & 0.68 & 0.69 & 0.69 \\* 
   &  &  40 & 0.60 & 0.65 & 0.70 & 0.71 & 0.72 & 0.72 & 0.72 & 0.73 & 0.72 & 0.72 & 0.72 & 0.74 & 0.73 & 0.73 \\* 
   &  &  50 & 0.64 & 0.69 & 0.72 & 0.74 & 0.74 & 0.74 & 0.74 & 0.74 & 0.74 & 0.74 & 0.74 & 0.76 & 0.75 & 0.75 \\* 
   &  &  60 & 0.68 & 0.72 & 0.74 & 0.75 & 0.75 & 0.75 & 0.75 & 0.75 & 0.75 & 0.75 & 0.75 & 0.78 & 0.76 & 0.76 \\* 
   &  &  70 & 0.73 & 0.75 & 0.77 & 0.77 & 0.77 & 0.77 & 0.77 & 0.77 & 0.77 & 0.76 & 0.76 & 0.81 & 0.78 & 0.78 \\* 
   &  &  80 & 0.76 & 0.76 & 0.77 & 0.77 & 0.77 & 0.77 & 0.77 & 0.77 & 0.77 & 0.77 & 0.77 & 0.81 & 0.79 & 0.79 \\* 
   &  &  90 & 0.78 & 0.78 & 0.78 & 0.78 & 0.78 & 0.78 & 0.78 & 0.77 & 0.77 & 0.77 & 0.77 & 0.82 & 0.80 & 0.79 \\* 
   &  & 100 & 0.80 & 0.79 & 0.79 & 0.79 & 0.79 & 0.78 & 0.78 & 0.78 & 0.78 & 0.78 & 0.78 & 0.83 & 0.81 & 0.80 \\* 
   &  & 200 & 0.85 & 0.83 & 0.82 & 0.81 & 0.81 & 0.80 & 0.80 & 0.80 & 0.79 & 0.79 & 0.79 & 0.86 & 0.84 & 0.83 \\* 
   &  & 500 & 0.87 & 0.85 & 0.83 & 0.82 & 0.81 & 0.81 & 0.80 & 0.80 & 0.80 & 0.80 & 0.80 & 0.87 & 0.86 & 0.85 \\* 
   &  & 1000 & 0.87 & 0.86 & 0.84 & 0.83 & 0.82 & 0.81 & 0.81 & 0.80 & 0.80 & 0.80 & 0.80 & 0.87 & 0.87 & 0.85 \\ 
   \hline
\multirow{12}{*}{0.5} & \multirow{12}{*}{12} &  20 & 0.68 & 0.72 & 0.74 & 0.74 & 0.75 & 0.75 & 0.75 & 0.75 & 0.75 & 0.75 & 0.74 & 0.78 & 0.78 & 0.78 \\* 
   &  &  30 & 0.76 & 0.78 & 0.79 & 0.79 & 0.79 & 0.79 & 0.79 & 0.79 & 0.79 & 0.79 & 0.78 & 0.83 & 0.83 & 0.83 \\* 
   &  &  40 & 0.80 & 0.81 & 0.80 & 0.80 & 0.80 & 0.79 & 0.79 & 0.79 & 0.79 & 0.79 & 0.78 & 0.85 & 0.85 & 0.84 \\* 
   &  &  50 & 0.82 & 0.81 & 0.81 & 0.80 & 0.80 & 0.79 & 0.79 & 0.79 & 0.79 & 0.78 & 0.78 & 0.86 & 0.85 & 0.85 \\* 
   &  &  60 & 0.84 & 0.82 & 0.82 & 0.81 & 0.81 & 0.80 & 0.80 & 0.80 & 0.79 & 0.79 & 0.79 & 0.86 & 0.86 & 0.85 \\* 
   &  &  70 & 0.85 & 0.83 & 0.82 & 0.81 & 0.81 & 0.80 & 0.80 & 0.80 & 0.79 & 0.79 & 0.79 & 0.86 & 0.86 & 0.86 \\* 
   &  &  80 & 0.85 & 0.83 & 0.82 & 0.81 & 0.81 & 0.81 & 0.80 & 0.80 & 0.80 & 0.79 & 0.79 & 0.87 & 0.86 & 0.86 \\* 
   &  &  90 & 0.85 & 0.83 & 0.82 & 0.81 & 0.81 & 0.80 & 0.80 & 0.80 & 0.80 & 0.79 & 0.79 & 0.87 & 0.86 & 0.86 \\* 
   &  & 100 & 0.86 & 0.84 & 0.83 & 0.82 & 0.81 & 0.81 & 0.81 & 0.80 & 0.80 & 0.80 & 0.80 & 0.87 & 0.87 & 0.86 \\* 
   &  & 200 & 0.87 & 0.85 & 0.84 & 0.83 & 0.82 & 0.81 & 0.81 & 0.81 & 0.80 & 0.80 & 0.80 & 0.87 & 0.87 & 0.87 \\* 
   &  & 500 & 0.87 & 0.85 & 0.84 & 0.83 & 0.82 & 0.82 & 0.81 & 0.81 & 0.81 & 0.80 & 0.80 & 0.87 & 0.87 & 0.87 \\* 
   &  & 1000 & 0.87 & 0.85 & 0.84 & 0.83 & 0.82 & 0.82 & 0.81 & 0.81 & 0.80 & 0.80 & 0.80 & 0.87 & 0.87 & 0.87 \\ 
   \hline
\multirow{12}{*}{0.5} & \multirow{12}{*}{100} &  20 & 0.56 & 0.59 & 0.62 & 0.64 & 0.65 & 0.66 & 0.66 & 0.66 & 0.66 & 0.66 & 0.66 & 0.64 & 0.66 & 0.65 \\* 
   &  &  30 & 0.59 & 0.63 & 0.67 & 0.69 & 0.70 & 0.71 & 0.71 & 0.71 & 0.71 & 0.71 & 0.71 & 0.70 & 0.71 & 0.72 \\* 
   &  &  40 & 0.64 & 0.69 & 0.72 & 0.73 & 0.74 & 0.74 & 0.74 & 0.74 & 0.74 & 0.74 & 0.74 & 0.75 & 0.75 & 0.75 \\* 
   &  &  50 & 0.69 & 0.72 & 0.74 & 0.74 & 0.75 & 0.75 & 0.75 & 0.75 & 0.75 & 0.75 & 0.75 & 0.78 & 0.76 & 0.77 \\* 
   &  &  60 & 0.73 & 0.74 & 0.76 & 0.76 & 0.77 & 0.77 & 0.77 & 0.77 & 0.77 & 0.77 & 0.76 & 0.80 & 0.79 & 0.79 \\* 
   &  &  70 & 0.77 & 0.76 & 0.77 & 0.77 & 0.77 & 0.77 & 0.77 & 0.77 & 0.77 & 0.77 & 0.77 & 0.82 & 0.80 & 0.80 \\* 
   &  &  80 & 0.79 & 0.77 & 0.78 & 0.78 & 0.78 & 0.78 & 0.78 & 0.77 & 0.77 & 0.77 & 0.77 & 0.83 & 0.81 & 0.80 \\* 
   &  &  90 & 0.81 & 0.78 & 0.78 & 0.78 & 0.78 & 0.78 & 0.78 & 0.78 & 0.78 & 0.78 & 0.78 & 0.83 & 0.82 & 0.81 \\* 
   &  & 100 & 0.82 & 0.79 & 0.79 & 0.79 & 0.79 & 0.79 & 0.78 & 0.78 & 0.78 & 0.78 & 0.78 & 0.84 & 0.82 & 0.82 \\* 
   &  & 200 & 0.85 & 0.83 & 0.82 & 0.81 & 0.80 & 0.80 & 0.80 & 0.80 & 0.79 & 0.79 & 0.79 & 0.86 & 0.85 & 0.84 \\* 
   &  & 500 & 0.87 & 0.84 & 0.83 & 0.82 & 0.81 & 0.81 & 0.80 & 0.80 & 0.80 & 0.80 & 0.80 & 0.87 & 0.87 & 0.86 \\* 
   &  & 1000 & 0.87 & 0.85 & 0.84 & 0.83 & 0.82 & 0.81 & 0.81 & 0.81 & 0.80 & 0.80 & 0.80 & 0.87 & 0.87 & 0.86 \\ 
   \hline
\multirow{12}{*}{0.5} & \multirow{12}{*}{200} &  20 & 0.54 & 0.56 & 0.58 & 0.60 & 0.61 & 0.62 & 0.62 & 0.63 & 0.63 & 0.63 & 0.63 & 0.60 & 0.62 & 0.62 \\* 
   &  &  30 & 0.56 & 0.61 & 0.64 & 0.66 & 0.67 & 0.68 & 0.68 & 0.68 & 0.69 & 0.69 & 0.68 & 0.67 & 0.69 & 0.69 \\* 
   &  &  40 & 0.60 & 0.65 & 0.69 & 0.70 & 0.71 & 0.72 & 0.72 & 0.72 & 0.72 & 0.72 & 0.72 & 0.72 & 0.72 & 0.72 \\* 
   &  &  50 & 0.63 & 0.68 & 0.71 & 0.73 & 0.73 & 0.74 & 0.74 & 0.74 & 0.74 & 0.74 & 0.74 & 0.75 & 0.74 & 0.74 \\* 
   &  &  60 & 0.69 & 0.71 & 0.73 & 0.74 & 0.74 & 0.74 & 0.74 & 0.74 & 0.74 & 0.74 & 0.74 & 0.77 & 0.75 & 0.76 \\* 
   &  &  70 & 0.73 & 0.74 & 0.75 & 0.76 & 0.76 & 0.76 & 0.76 & 0.76 & 0.76 & 0.76 & 0.76 & 0.80 & 0.78 & 0.78 \\* 
   &  &  80 & 0.76 & 0.75 & 0.76 & 0.76 & 0.77 & 0.77 & 0.77 & 0.77 & 0.76 & 0.76 & 0.76 & 0.81 & 0.79 & 0.79 \\* 
   &  &  90 & 0.78 & 0.76 & 0.77 & 0.77 & 0.77 & 0.77 & 0.77 & 0.77 & 0.77 & 0.77 & 0.77 & 0.82 & 0.79 & 0.79 \\* 
   &  & 100 & 0.80 & 0.77 & 0.78 & 0.78 & 0.78 & 0.78 & 0.78 & 0.78 & 0.78 & 0.78 & 0.78 & 0.83 & 0.80 & 0.80 \\* 
   &  & 200 & 0.85 & 0.82 & 0.81 & 0.81 & 0.80 & 0.80 & 0.80 & 0.79 & 0.79 & 0.79 & 0.79 & 0.86 & 0.84 & 0.83 \\* 
   &  & 500 & 0.87 & 0.84 & 0.83 & 0.82 & 0.81 & 0.81 & 0.80 & 0.80 & 0.80 & 0.80 & 0.80 & 0.87 & 0.86 & 0.85 \\* 
   &  & 1000 & 0.87 & 0.85 & 0.83 & 0.82 & 0.82 & 0.81 & 0.81 & 0.80 & 0.80 & 0.80 & 0.80 & 0.87 & 0.87 & 0.86 \\ 
   \hline
\multirow{12}{*}{1} & \multirow{12}{*}{12} &  20 & 0.68 & 0.71 & 0.73 & 0.74 & 0.74 & 0.75 & 0.75 & 0.74 & 0.75 & 0.74 & 0.74 & 0.75 & 0.75 & 0.76 \\* 
   &  &  30 & 0.75 & 0.77 & 0.78 & 0.78 & 0.79 & 0.78 & 0.79 & 0.78 & 0.78 & 0.78 & 0.78 & 0.81 & 0.81 & 0.82 \\* 
   &  &  40 & 0.80 & 0.79 & 0.79 & 0.79 & 0.79 & 0.79 & 0.79 & 0.79 & 0.79 & 0.79 & 0.78 & 0.83 & 0.83 & 0.84 \\* 
   &  &  50 & 0.82 & 0.80 & 0.80 & 0.80 & 0.79 & 0.79 & 0.79 & 0.79 & 0.79 & 0.78 & 0.78 & 0.85 & 0.85 & 0.85 \\* 
   &  &  60 & 0.84 & 0.80 & 0.80 & 0.80 & 0.80 & 0.80 & 0.80 & 0.79 & 0.79 & 0.79 & 0.79 & 0.86 & 0.85 & 0.86 \\* 
   &  &  70 & 0.85 & 0.81 & 0.81 & 0.80 & 0.80 & 0.80 & 0.80 & 0.80 & 0.79 & 0.79 & 0.79 & 0.86 & 0.86 & 0.86 \\* 
   &  &  80 & 0.85 & 0.81 & 0.81 & 0.81 & 0.80 & 0.80 & 0.80 & 0.80 & 0.80 & 0.79 & 0.79 & 0.86 & 0.86 & 0.86 \\* 
   &  &  90 & 0.85 & 0.82 & 0.81 & 0.81 & 0.80 & 0.80 & 0.80 & 0.80 & 0.79 & 0.79 & 0.79 & 0.86 & 0.86 & 0.86 \\* 
   &  & 100 & 0.86 & 0.82 & 0.82 & 0.81 & 0.81 & 0.81 & 0.80 & 0.80 & 0.80 & 0.80 & 0.80 & 0.87 & 0.87 & 0.87 \\* 
   &  & 200 & 0.87 & 0.83 & 0.82 & 0.82 & 0.81 & 0.81 & 0.81 & 0.81 & 0.80 & 0.80 & 0.80 & 0.87 & 0.87 & 0.87 \\* 
   &  & 500 & 0.87 & 0.83 & 0.82 & 0.82 & 0.82 & 0.81 & 0.81 & 0.81 & 0.80 & 0.80 & 0.80 & 0.87 & 0.87 & 0.87 \\* 
   &  & 1000 & 0.87 & 0.83 & 0.82 & 0.82 & 0.81 & 0.81 & 0.81 & 0.81 & 0.80 & 0.80 & 0.80 & 0.87 & 0.87 & 0.87 \\ 
   \hline
\multirow{12}{*}{1} & \multirow{12}{*}{100} &  20 & 0.56 & 0.58 & 0.61 & 0.62 & 0.64 & 0.64 & 0.64 & 0.65 & 0.65 & 0.66 & 0.65 & 0.61 & 0.65 & 0.61 \\* 
   &  &  30 & 0.59 & 0.60 & 0.64 & 0.67 & 0.68 & 0.68 & 0.69 & 0.69 & 0.69 & 0.69 & 0.69 & 0.65 & 0.69 & 0.65 \\* 
   &  &  40 & 0.64 & 0.65 & 0.69 & 0.71 & 0.72 & 0.72 & 0.72 & 0.72 & 0.72 & 0.72 & 0.72 & 0.70 & 0.71 & 0.71 \\* 
   &  &  50 & 0.69 & 0.68 & 0.71 & 0.72 & 0.72 & 0.72 & 0.73 & 0.73 & 0.73 & 0.73 & 0.73 & 0.72 & 0.72 & 0.72 \\* 
   &  &  60 & 0.73 & 0.70 & 0.73 & 0.74 & 0.74 & 0.75 & 0.75 & 0.75 & 0.75 & 0.75 & 0.75 & 0.75 & 0.74 & 0.75 \\* 
   &  &  70 & 0.77 & 0.72 & 0.74 & 0.75 & 0.75 & 0.75 & 0.75 & 0.75 & 0.75 & 0.75 & 0.75 & 0.76 & 0.75 & 0.76 \\* 
   &  &  80 & 0.79 & 0.73 & 0.74 & 0.75 & 0.76 & 0.76 & 0.76 & 0.76 & 0.76 & 0.76 & 0.76 & 0.78 & 0.76 & 0.78 \\* 
   &  &  90 & 0.81 & 0.74 & 0.75 & 0.76 & 0.76 & 0.76 & 0.76 & 0.77 & 0.76 & 0.76 & 0.76 & 0.79 & 0.77 & 0.79 \\* 
   &  & 100 & 0.82 & 0.75 & 0.76 & 0.77 & 0.77 & 0.77 & 0.77 & 0.77 & 0.77 & 0.77 & 0.77 & 0.80 & 0.78 & 0.80 \\* 
   &  & 200 & 0.85 & 0.80 & 0.80 & 0.80 & 0.79 & 0.79 & 0.79 & 0.79 & 0.79 & 0.79 & 0.79 & 0.85 & 0.83 & 0.84 \\* 
   &  & 500 & 0.87 & 0.82 & 0.81 & 0.81 & 0.81 & 0.80 & 0.80 & 0.80 & 0.80 & 0.80 & 0.80 & 0.87 & 0.87 & 0.86 \\* 
   &  & 1000 & 0.87 & 0.83 & 0.82 & 0.82 & 0.81 & 0.81 & 0.81 & 0.80 & 0.80 & 0.80 & 0.80 & 0.87 & 0.87 & 0.87 \\ 
   \hline
\multirow{12}{*}{1} & \multirow{12}{*}{200} &  20 & 0.54 & 0.55 & 0.57 & 0.59 & 0.60 & 0.61 & 0.61 & 0.62 & 0.62 & 0.62 & 0.62 & 0.57 & 0.61 & 0.58 \\* 
   &  &  30 & 0.57 & 0.58 & 0.62 & 0.65 & 0.66 & 0.67 & 0.67 & 0.67 & 0.67 & 0.67 & 0.67 & 0.63 & 0.67 & 0.64 \\* 
   &  &  40 & 0.60 & 0.62 & 0.66 & 0.68 & 0.69 & 0.70 & 0.70 & 0.70 & 0.70 & 0.70 & 0.71 & 0.67 & 0.70 & 0.68 \\* 
   &  &  50 & 0.64 & 0.65 & 0.69 & 0.70 & 0.71 & 0.72 & 0.72 & 0.72 & 0.72 & 0.72 & 0.72 & 0.70 & 0.72 & 0.70 \\* 
   &  &  60 & 0.68 & 0.67 & 0.70 & 0.71 & 0.72 & 0.72 & 0.72 & 0.72 & 0.72 & 0.72 & 0.72 & 0.71 & 0.72 & 0.71 \\* 
   &  &  70 & 0.73 & 0.69 & 0.72 & 0.73 & 0.73 & 0.74 & 0.74 & 0.74 & 0.74 & 0.74 & 0.74 & 0.74 & 0.73 & 0.74 \\* 
   &  &  80 & 0.76 & 0.71 & 0.73 & 0.74 & 0.74 & 0.74 & 0.74 & 0.74 & 0.74 & 0.74 & 0.74 & 0.75 & 0.74 & 0.75 \\* 
   &  &  90 & 0.78 & 0.72 & 0.73 & 0.74 & 0.74 & 0.75 & 0.75 & 0.75 & 0.75 & 0.75 & 0.75 & 0.76 & 0.75 & 0.76 \\* 
   &  & 100 & 0.80 & 0.73 & 0.74 & 0.75 & 0.75 & 0.76 & 0.76 & 0.76 & 0.76 & 0.76 & 0.76 & 0.78 & 0.76 & 0.77 \\* 
   &  & 200 & 0.85 & 0.78 & 0.79 & 0.79 & 0.79 & 0.79 & 0.79 & 0.79 & 0.79 & 0.79 & 0.79 & 0.84 & 0.81 & 0.82 \\* 
   &  & 500 & 0.87 & 0.82 & 0.81 & 0.81 & 0.80 & 0.80 & 0.80 & 0.80 & 0.80 & 0.80 & 0.79 & 0.86 & 0.85 & 0.85 \\* 
   &  & 1000 & 0.87 & 0.83 & 0.82 & 0.81 & 0.81 & 0.81 & 0.80 & 0.80 & 0.80 & 0.80 & 0.80 & 0.87 & 0.87 & 0.87 \\ 
\caption{Mean C-index, averaged across all test data sets and subgroups, for all 252 simulation scenarios and all 14 Cox model types} 
\label{tab_sim_CI}
\end{longtable}
\end{landscape}
\clearpage

\end{document}